\documentclass{article}

\pdfoutput=1  % Ensures pdflatex processing by arXiv

\usepackage{authblk}
\usepackage{amsmath}
\usepackage{graphicx}

\begin{document}

%\runningheads{I.~Glendinning et al.}{Identification of an interferer by
%                                     comparison with known carriers}

\title{Identification of the source of an interferer
       by comparison with known carriers using a single satellite}

%\title{Identification of an interfering transmitter from a set of known
%       transmitters using a single satellite}

%\title{Identifying the source of an unknown signal from a set of known
%       transmitters using a single satellite}

\author[1]{Ian~Glendinning\thanks{Correspondence to: AIT Austrian Institute
           of Technology GmbH, Giefinggasse 4, 1210 Vienna, Austria.
           E-mail: ian.glendinning@ait.ac.at}}
\author[1]{Michael~N\"olle}
\author[2]{Christian Hausleitner}
\author[2]{Erwin Greilinger}

\affil[1]{AIT Austrian Institute of Technology GmbH,
         Giefinggasse 4, 1210 Vienna, Austria}
\affil[2]{Atos Convergence Creators GmbH,
         Autokaderstra\ss e 29, 1210 Vienna, Austria}

% \author{Ian~Glendinning\affil{1}\corrauth, Michael~N\"olle\affil{1},
%         Christian Hausleitner\affil{2} and Erwin Greilinger\affil{2}}
% 
% \address{\affilnum{1}AIT Austrian Institute of Technology GmbH,
%          Giefinggasse 4, 1210 Vienna, Austria\break
%          \affilnum{2}Atos Convergence Creators GmbH,
%          Autokaderstra\ss e 29, 1210 Vienna, Austria}

%\corraddr{AIT Austrian Institute of Technology GmbH, Giefinggasse 4,
%1210 Vienna, Austria. E-mail: ian.glendinning@ait.ac.at}

%\cgsn{\"Osterreichische Forschungsf\"orderungsgesellschaft FFG}{833427}

\maketitle

\begin{abstract}
We describe a method for identifying the source of a satellite interferer
using a single satellite. The technique relies on the fact that the strength
of a carrier signal measured at the downlink station varies with time due to
a number of factors, and we use a quantum-inspired algorithm to compute a
`signature' for a signal, which captures part of the pattern of variation that
is characteristic of the uplink antenna. We define a distance measure to
numerically quantify the degree of similarity between two signatures, and by
computing the distances between the signature for an interfering carrier and
the signatures of the known carriers being relayed by the same satellite at
the same time, we can identify the antenna that the interferer originated
from, if a known carrier is being relayed from it. As a proof of concept we
evaluate the performance of the technique using a simple statistical model
applied to measured carrier data.

\end{abstract}

%\keywords{interferer; identification; single satellite;
%          quantum inspired; singular value decomposition}

%\maketitle

\section{Introduction}

The increasing demand for satellite communication links has led to an
increasing number of satellite signals, and to an increasing amount of uplink
interference. The causes of this interference include the growth in the number
of small ground terminals, low quality equipment, poor installations and
maintenance, uplink personnel mistakes (human error), faulty equipment,
incorrectly pointed antennas, adjacent satellite interference, terrestrial
service interference, and sometimes intentional jamming~\cite{Coleman2014,Oez2013}.
Satellite operators are therefore increasingly interested in solutions not
only for detecting interference, which is the main task of a satellite
monitoring system, but also to identify its source.

The traditional approach
is to geographically localize, or geolocate, the transmitting station of an
interferer. However, most localization systems need to receive the
interference signal via two adjacent satellites in order to
perform geolocation~\cite{Chestnut1982,SmithSteffes1989,EfflandGSW1991,
HaworthSBC1997,GrantSD2013}, and there are a number of limitations
associated with this approach:
\begin{itemize}
\item An adjacent satellite must be available that is equipped with
      transponder(s) receiving components of the interfering signal and a
      reference signal (same uplink frequency range, same polarization).
\item The interference and reference signals need to have enough crosstalk
      energy between the primary and adjacent satellites to achieve a
      sufficient level of correlation.
\item Accurate ephemeris data must be available for both satellites.
\item The reference signal needs to be received from both satellites via
      transponders operating with the same physical local oscillators (LOs)
      as the transponders re-transmitting the interference signal.
\item If the system is installed at only one earth station, the downlink
      signals of both satellites need to be receivable at this earth station
      (downlink beams of both satellites need to cover the measurement site
      location). If this is not possible (beams pointing to different
      locations) the system needs to be installed at different locations
      inside the beams.
\item A region is identified in which the transmitter is likely to reside, but
      additional steps are often necessary to actually identify the
      transmitter.
\end{itemize}
Geolocation can also be performed using crosstalk measurements between signals
received from multiple antennas/beams belonging to the same
satellite~\cite{Adeogun2013,Fredrick2014}, but this approach has the drawback
that additional payload resources are needed (antennas, transponders) or that
operations must be interrupted to release resources. It has also been shown
that frequency measurements of signals from a single satellite, taken at
different times, can be used to locate an unknown
emitter~\cite{HoCD2013,AshkanMCO2016}, but this approach is extremely
susceptible to frequency instability introduced by the uplink terminal, which
leads to very high localization errors unless the terminal's frequency
stability is better than $\pm 1 \times 10^{-12}$ per day, which can be achieved
for example via synchronization with a Galileo/GPS/GNNS disciplined frequency
reference oscillator.

Here we describe a method able to identify the source of an interferer using
a single satellite, based on the variation of signal strength with time,
measured at the downlink station. It is a variant on a method that is
the subject of Austrian patent~\cite{Noelle2015a} and of international,
US, and European patent applications~\cite{Noelle2015b,Noelle2015c,Noelle2015d}.
The main benefit of our approach is that it enables identification of unknown
RFI transmitters based only on measurements of power variations. This overcomes
the constraints of the above methods. Even in the case that the position of
the transmitter of a `matching' reference signal is not known, the result can
be used for resolving the interference case by contacting the satellite
operator's accounting department to get in touch with the customer (the
individual operating the uplink) who is potentially causing the interference.

The main limitation of the approach is that the interferer must be from a known 
antenna from which a known carrier is also being relayed. This means that the 
method only works for antennas that are transmitting at least two carriers, and 
that the interferer must be from a known antenna. As an example, in 2018 
roughly 30\% of antennas pointing to a big satellite fleet transmitted two or 
more carriers, and in 2012 T\"urksat reported that just 3\% of interference was 
due to unknown carriers~\cite{Oez2013}, so our method is applicable in a 
significant number of cases. It is not a substitute for the traditional 
geolocation approach with adjacent satellites (based on TDOA/FDOA 
measurements), but in the case that the traditional approach does not work (no 
adjacent satellite available; different beam coverage; not the same uplink 
frequency; etc.), which happens more than 60\% to 70\% of the time, it offers an 
additional possibility.

The rest of this paper is structured as follows. Section~\ref{our-method}
outlines our method and discusses the power variations that it relies on,
and the limitations of the approach, as well its quantum-inspired aspects.
In section~\ref{compute-signature} we explain in detail how to compute the
signature, and in section~\ref{quantify-similarity} how to quantify the
similarity between signatures. Section~\ref{performance-evaluation} analyses
the performance of the method, and section \ref{conclusion} concludes.

\section{Method}
\label{our-method}

Our method relies on the fact that the signal strength of a carrier that is
measured at the downlink station varies with time due to a number of factors,
and the technique is capable of identifying the antenna that an interferer
originated from if another `known' carrier is being relayed by the
satellite at the same time from the same antenna.
It turns out that there are similarities in the patterns of variation of
signal strength for carriers originating from the same uplink antenna, and we
found we were able to compute a `signature' for the variation of signal
strength for each carrier, that captures part of the pattern of variation
that is characteristic of the uplink antenna.

In order to numerically quantify the degree of similarity or difference
between two signatures, we compute a `distance' between them, which is a
number between 0 and 1. If the distance is close to zero, the signatures are
similar (if they are identical the distance is zero), and if it is close to
one, they are very different. This distance between signatures turns out
to be lower on average for carriers from the same antenna than for carriers
from different antennas, and by comparing the signature for an interfering
carrier with the signatures for the other carriers being relayed by the
same satellite, we can rank them according to their degree of similarity.

The causes of power variations in a received carrier include:
\begin{itemize}
\item Power variations from signal-sending hardware (satellite
      modem, frequency converter, power amplifier, etc.)
\item Satellite movement versus antenna pattern and pointing mechanism
      (antenna tracking the satellite position or constant bearing towards the
      satellite, antenna pointing variations due to wind)
\item Atmospheric losses due gases and hydrometeors
\item Faraday rotation in the ionosphere
\item Noise contributions (terrestrial noise picked up from the surface of the
      earth, receiver noise in both satellite and Rx ground station,
      atmospheric noise, extra-terrestrial noise from the sun and moon, etc. )
\end{itemize}

Our approach only works if the signal power is not strongly affected by the
satellite transponders, so it applies to transparent transponders working with
constant gain (fixed gain mode) and which are not saturated. In the case of
saturation and/or if Automatic Level Control (ALC) is applied, the sensitivity
of the  measurements can be severely reduced, requiring different measurement
settings (high averaging) and a reference carrier that is affected by the
same mode of transponder operation in order for the approach to work. It does
not work with regenerative transponders.
The method works well with different transponders. A small reduction
in similarity is introduced by frequency dependency, meaning that if
a carrier's frequency is different by e.g. 1 GHz (Ku-Band) the level of
similarity in power fluctuation is slightly reduced. More degradation of level
of similarity comes from different polarization, but the similarity is still
high enough for successful detection.

The sensitivity of the measurement could perhaps be increased if downlink path
influences, such as the power variation of a beacon signal and/or the
transponder noise floor and/or the average power variation of all the
signals on the downlink, were subtracted from the unknown signal and the known
signal.

\subsection{A quantum-inspired algorithm}

The algorithm we present here to calculate the similarity between two
carriers is based on the one described in the patent
applications~\cite{Noelle2015a,Noelle2015b,Noelle2015c,Noelle2015d},
which is a so-called `quantum inspired algorithm', in which concepts from
quantum information theory are applied to the representation and processing
of classical information~\cite{Noelle2007,Noelle2011,NoelleSB2013}.
The first step in developing a quantum inspired algorithm is to find a
suitable encoding of the information as quantum states, which can then be
manipulated using the well-developed mathematical techniques of quantum
information theory. In the patent applications the signal was
encoded in terms of {\em qubits} (quantum bits), which has advantages
when the absolute value of the signal contains significant information,
but in this case we subtract a running average from it, so there is no
advantage in using the qubit encoding. In the qubit representation each
signal value is mapped to
two values in a vector that is normalized to have an $\ell_2$ norm
(Euclidean norm) of $1$, but here we map each signal value to a single
value in a normalized vector. Both kinds of vector are valid
representations of quantum states in a
Hilbert space\footnote{Although here we are working with vectors in a real
inner product state, which is a special case of a Hilbert space, this
approach can be generalized to use complex vectors, as explained
in~\cite{Noelle2015a,Noelle2015b,Noelle2015c,Noelle2015d}}. In the patent
application we used the Schmidt decomposition~\cite{BogdanovBLFC2016,
NielsenC2011} to analyse the 24-hour periodic structure of the signal,
and to extract its principal components~\cite{Jolliffe2002}, one of
which served as a `signature', and we defined a natural distance measure
in terms of the $\ell_2$ norm. Here we use the singular value decomposition,
which is  equivalent to the Schmidt decomposition, together with the
same distance measure.

Although the algorithm presented here was inspired by concepts from quantum
information theory, we have expressed it in terms of a finite-dimensional
inner product space and the singular value decomposition, which are familiar
concepts in statistical signal processing.

\section{Computing the Signature}
\label{compute-signature}

The signature is computed from a sequence of satellite downlink
$\mathrm{EIRP}$ (Equivalent Isotropically Radiated Power) values
representing the variation from the uplink signal, measured in
$\mathrm{dBW}$, calculated as follows:

\begin{equation}
\mathrm{EIRP [dBW] = P_{sa} [dBm] + L_{fs} [dB] - G_{ant} [dB]
                     - G_{path} [dB] - 30 dB} ,
\end{equation}
where $\mathrm{P_{sa}}$ is the power at the input of the monitoring device
(Spectrum Analyzer), $\mathrm{L_{fs}}$ is free space loss, $\mathrm{G_{ant}}$
is the receiving antenna gain, $\mathrm{G_{path}}$ is the path gain from the
antenna feed to the spectrum analyzer, and $\mathrm{30 dB}$ is the conversion
from $\mathrm{dBm}$ to $\mathrm{dBW}$.
The measurement process takes into account the contribution of noise when
calculating $\mathrm{P_{sa}}$, subtracting it from the received signal
(power + noise).  The SNR limitation of this process is about 3 dB, meaning
that signals with SNR  below 3 dB are not taken into account. This limit is
chosen in order to keep the additional error due to estimation and subtraction
of noise small. If noise was not subtracted, the measurement would suffer
from sensitivity in terms of reduced amplitude of power fluctuations

The $\mathrm{EIRP}$ values must be
equally spaced in time, so if the raw data was not measured at a fixed time
interval, it must be interpolated to give values that are equally spaced in
time. The data used for the results presented in this paper was interpolated
at three minute intervals, which was roughly half the average interval between
measurements in the raw data.
For the calculation of the signature it is the variation of the
$\mathrm{EIRP}$ with time is important rather than its absolute value, so the
absolute value is removed in following steps.

\subsection{Expressing EIRP values as a state vector}
\label{state-vector}

Given a vector of $\mathrm{EIRP}$ values $(EIRP_1, EIRP_2, \dots ,EIRP_i,
\dots ,EIRP_N)$ corresponding to times $t_1, t_2, \dots ,t_i, \dots ,t_N$,
with equal intervals between them, we first subtract a
running average from the $\mathrm{EIRP}$ values,
using a window of a specified width in time. This has the effect that
constant differences between the average $\mathrm{EIRP}$ values of two
carriers, which are not relevant to the signature, are not taken into
account. To generate the results presented here, we used a Gaussian window
with a standard deviation of 6 hours, meaning that average differences
between one day and the next were also removed, and we
call the resulting values ${\mathbf E} = (E_1, E_2, \dots ,E_i, \dots ,E_N)$.

Now, for our quantum-inspired algorithm we wish to encode these values as
a quantum state vector, which we call
\begin{equation}
{\mathbf q}=(q_1, q_2, \dots ,q_i, \dots q_N) .
\label{q-def}
\end{equation}
To qualify as a state vector $\mathbf q$ must have a norm of
one~\cite{NielsenC2011}, i.e. $\|{\mathbf q}\| = 1$, where
for the special case of a quantum state for which all of the $q_i$ are real
numbers the norm is defined as\footnote{For a general quantum state the
$q_i$ are complex numbers and the norm is defined as
$\|{\mathbf q}\| = \sqrt{{\mathbf q}^\dagger {\mathbf q}}$}
\begin{equation}
\|{\mathbf q}\| = \sqrt{{\mathbf q}^\textrm{T} {\mathbf q}} .
\label{norm-def}
\end{equation}
This means that ${\mathbf q}^\textrm{T} {\mathbf q} = \sum_i q_i^2 = 1$,
and the $q_i^2$ values can be interpreted as probabilities because they
are between 0 and 1 and their sum is one.
In quantum mechanics the $q_i$ values are known as (probability)
amplitudes and the $q_i^2$ values correspond to the probabilities of
particular outcomes of measurements. It is the amplitudes that are the
fundamental quantities, so we choose to encode the signal in
terms of them. We first define
\begin{equation}
e_i = \frac{E_i + E_{max}}{2 E_{max}} ,
\label{ei-def}
\end{equation}
where $E_{max}$ is the maximum absolute value of $E_i$ for all $i$, that is,
$E_{max} = \max_i |E_i|$, so that $0 \leq e_i \leq 1$.
We then let\footnote{We could have defined $p_i$ directly in terms of
$E_i + E_{max}$, but we find it clearer to introduce $e_i$ as an
intermediary}
\begin{equation}
p_i = \frac{e_i}{\sum_j e_j} ,
\label{pi-def}
\end{equation}
so that the $p_i$ values form a valid probability distribution, with
$0 \le p_i \le 1$ and $\sum_i p_i = 1$, and we derive our $q_i$ values
from the $p_i$ values, by defining $q_i^2 = p_i$, so that
\begin{equation}
q_i = \sqrt{p_i} ,
\label{qi-def}
\end{equation}
taking the positive square root, so that $0 \le q_i \le 1$.

It might be thought that it would be simpler to encode the data as a
state vector by defining ${\mathbf e}=(e_1, e_2, \dots ,e_i, \dots e_N)$ and
${\mathbf r} = {\mathbf e} / \|{\mathbf e}\|$, thus avoiding the square root
in equation~\ref{qi-def}, but although ${\mathbf r}$ is a valid state vector,
we found that the `quantum-inspired' approach of
encoding the data as probability amplitudes in ${\mathbf q}$ gave slightly
better results in terms of being able to distinguish between pairs of carriers
from the same antenna and from different antennas when using the distance
measure defined in section~\ref{quantify-similarity}.

\subsection{Generating `eigensignals' for a state vector}

\begin{figure}[htb]
\centering
\includegraphics[width=71mm]{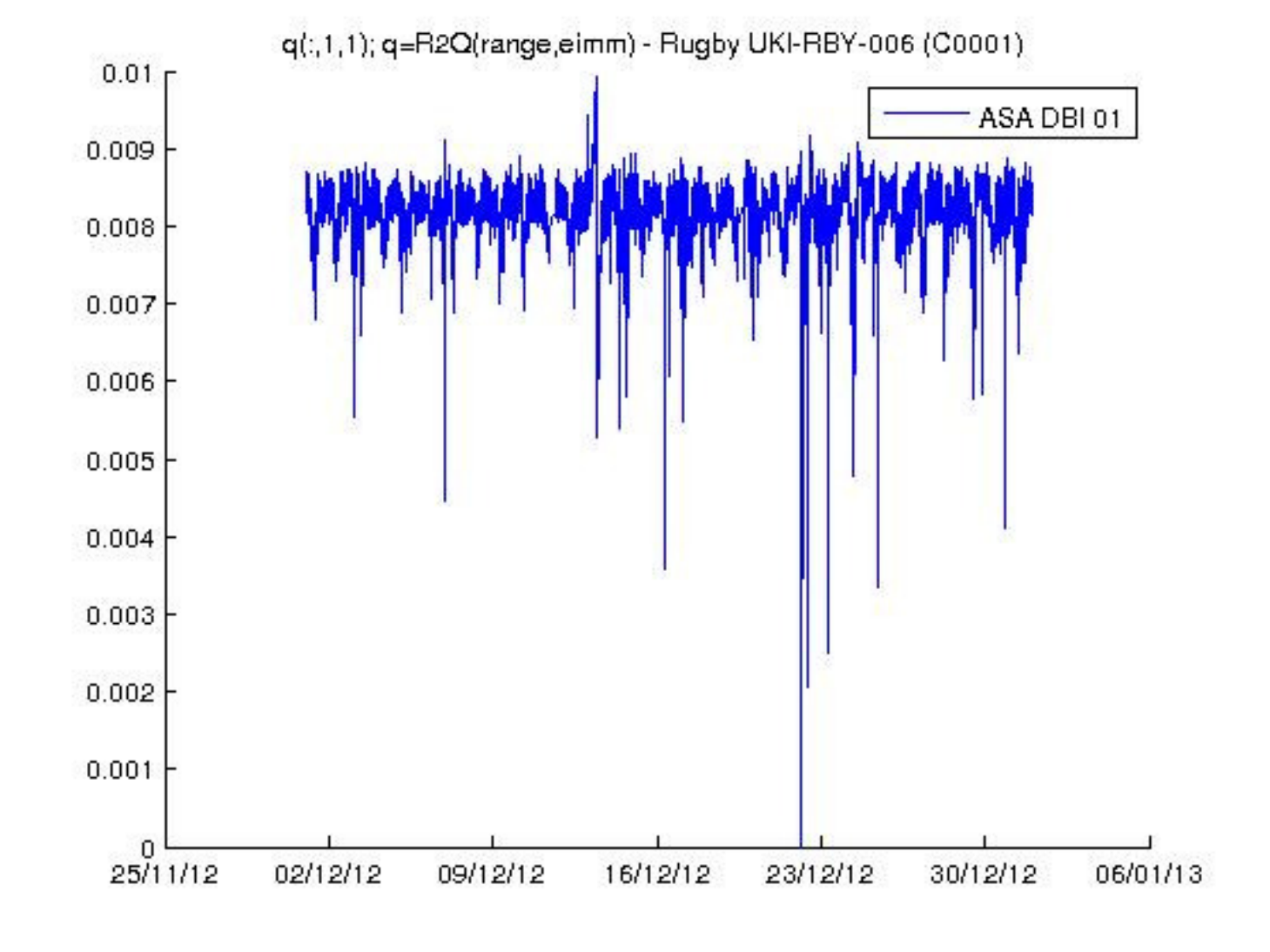}
\includegraphics[width=71mm]{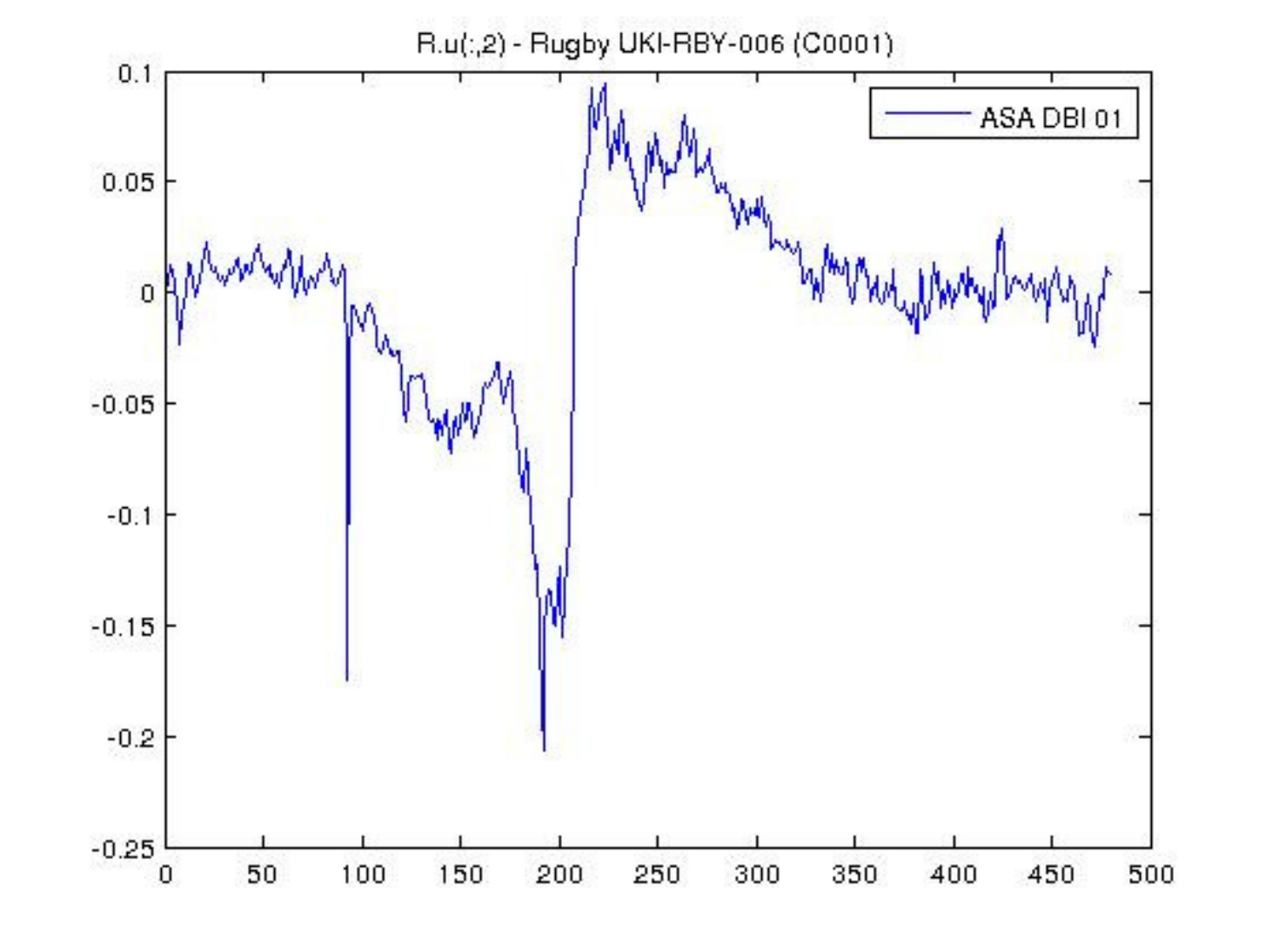}
\caption{$q_i$ vs $i$ (left) and ${\bf v}_2$ (right), for a signal
from Rugby}
\label{rugby1}
\end{figure}

\begin{figure}[htb]
\centering
\includegraphics[width=71mm]{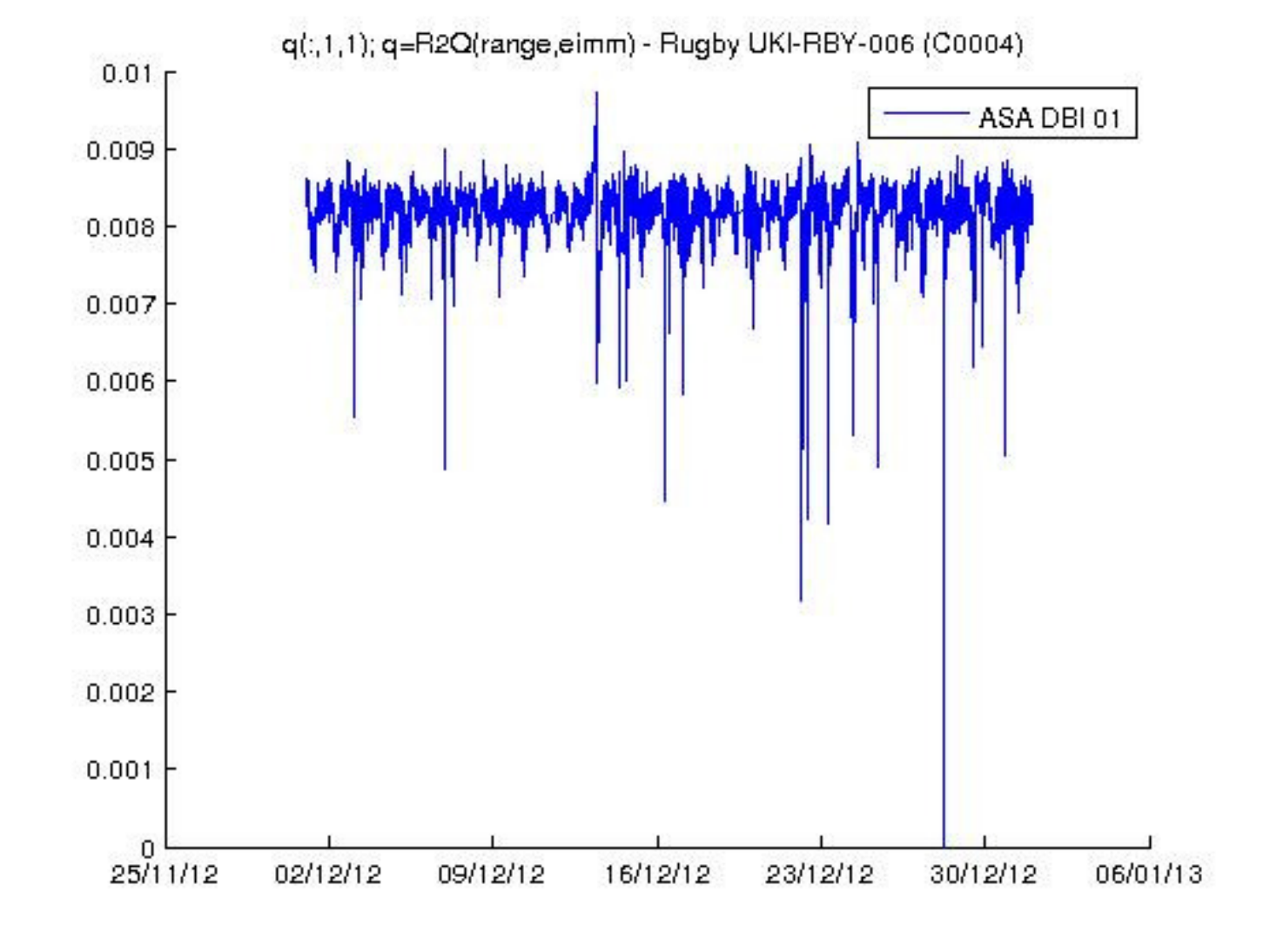}
\includegraphics[width=71mm]{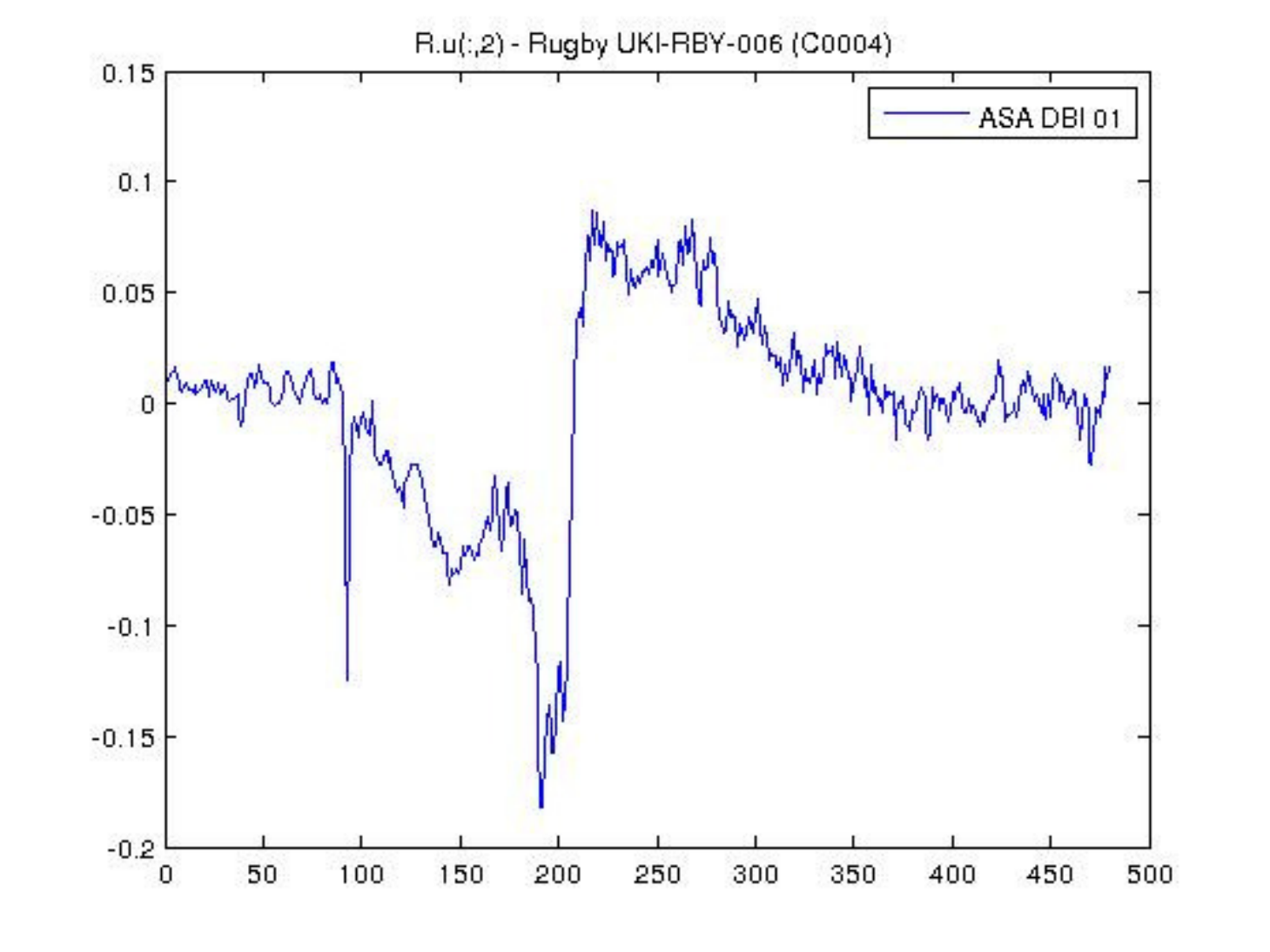}
\caption{$q_i$ vs $i$ (left) and ${\bf v}_2$ (right),
for a second signal from Rugby}
\label{rugby2}
\end{figure}

\begin{figure}[htb]
\centering
\includegraphics[width=71mm]{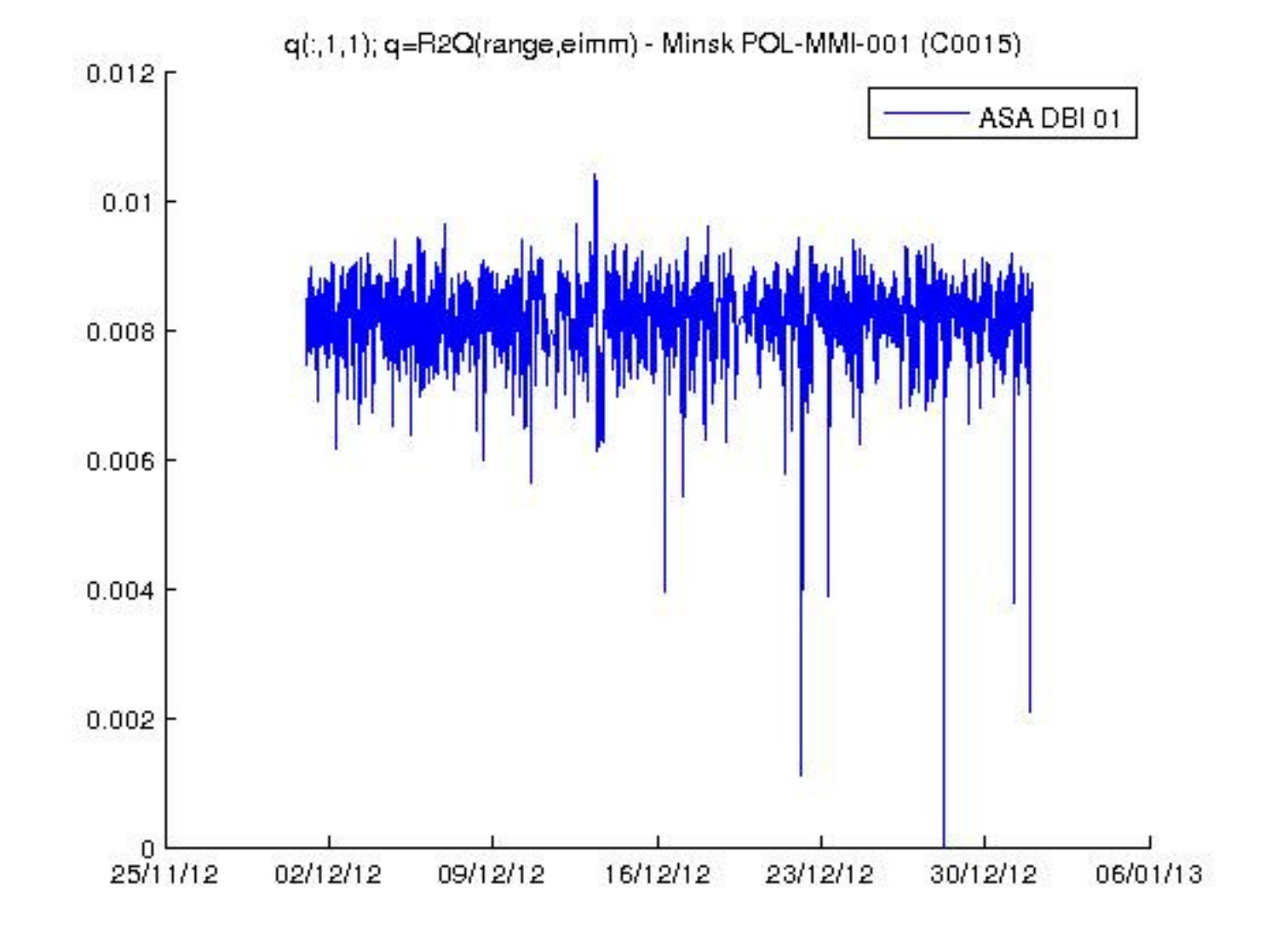}
\includegraphics[width=71mm]{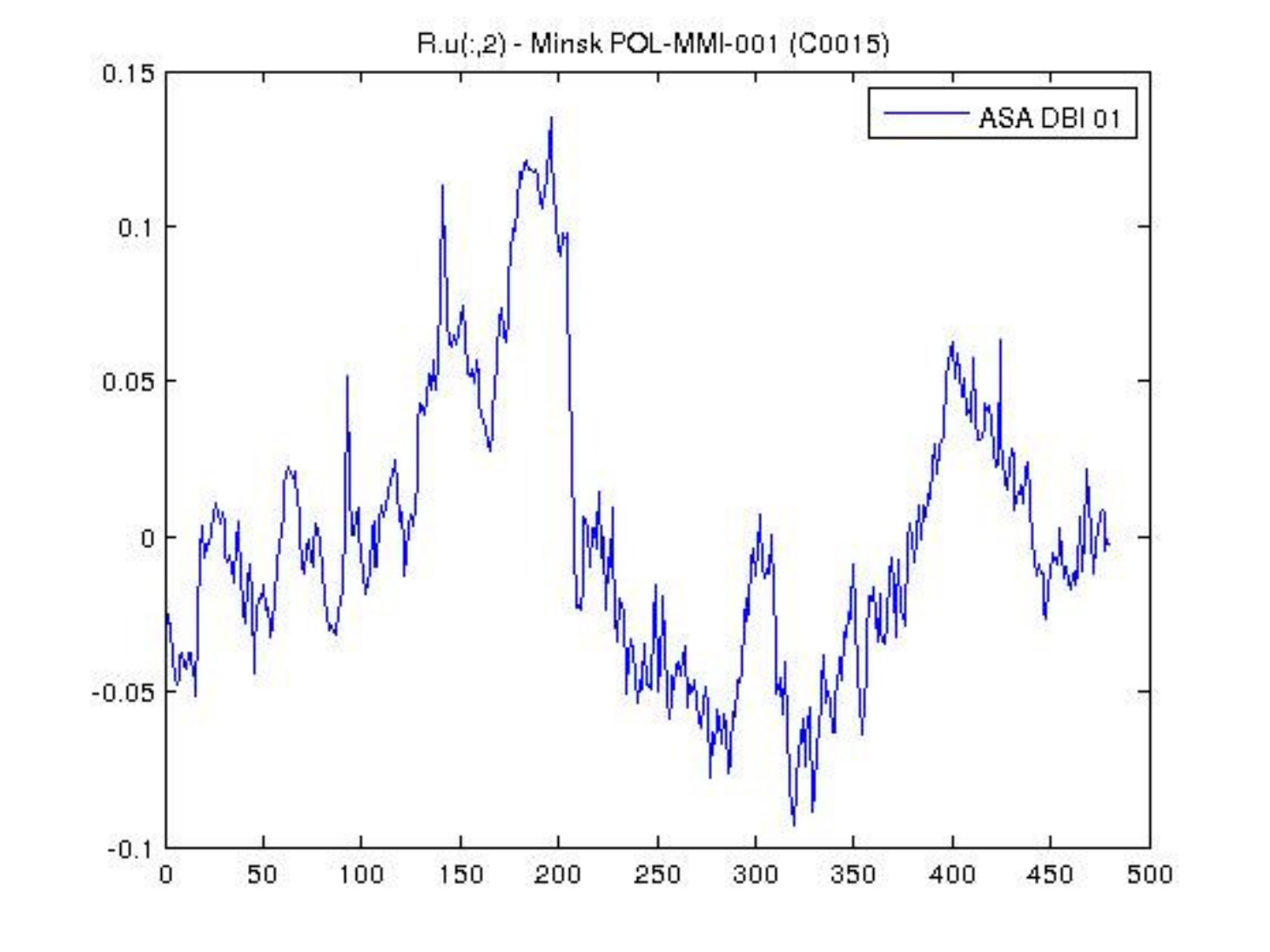}
\caption{$q_i$ vs $i$ (left) and ${\bf v}_2$ (right),
for a signal from Mi\'{n}sk Mazowiecki}
\label{minsk1}
\end{figure}

\begin{figure}[htb]
\centering
\includegraphics[width=71mm]{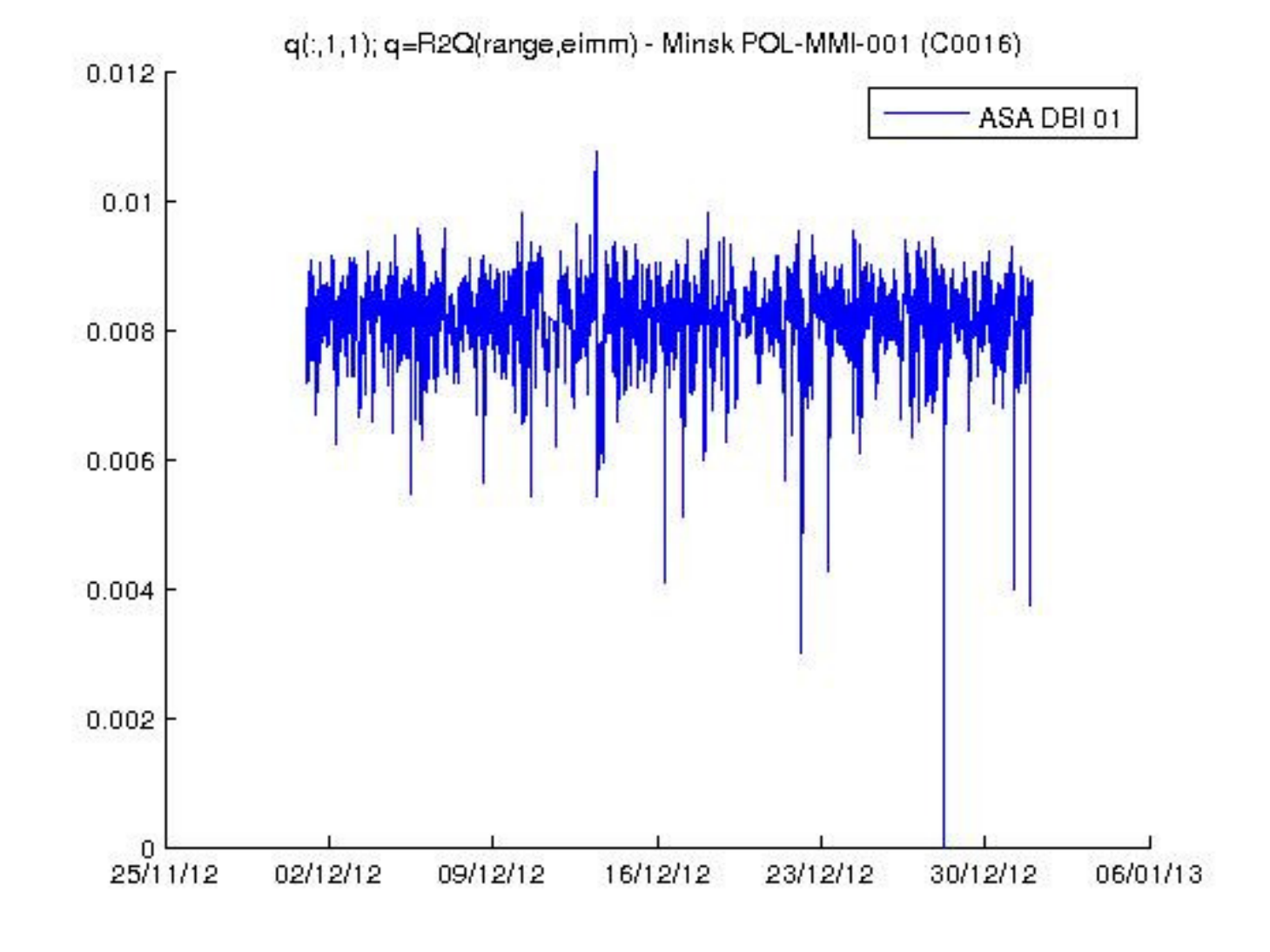}
\includegraphics[width=71mm]{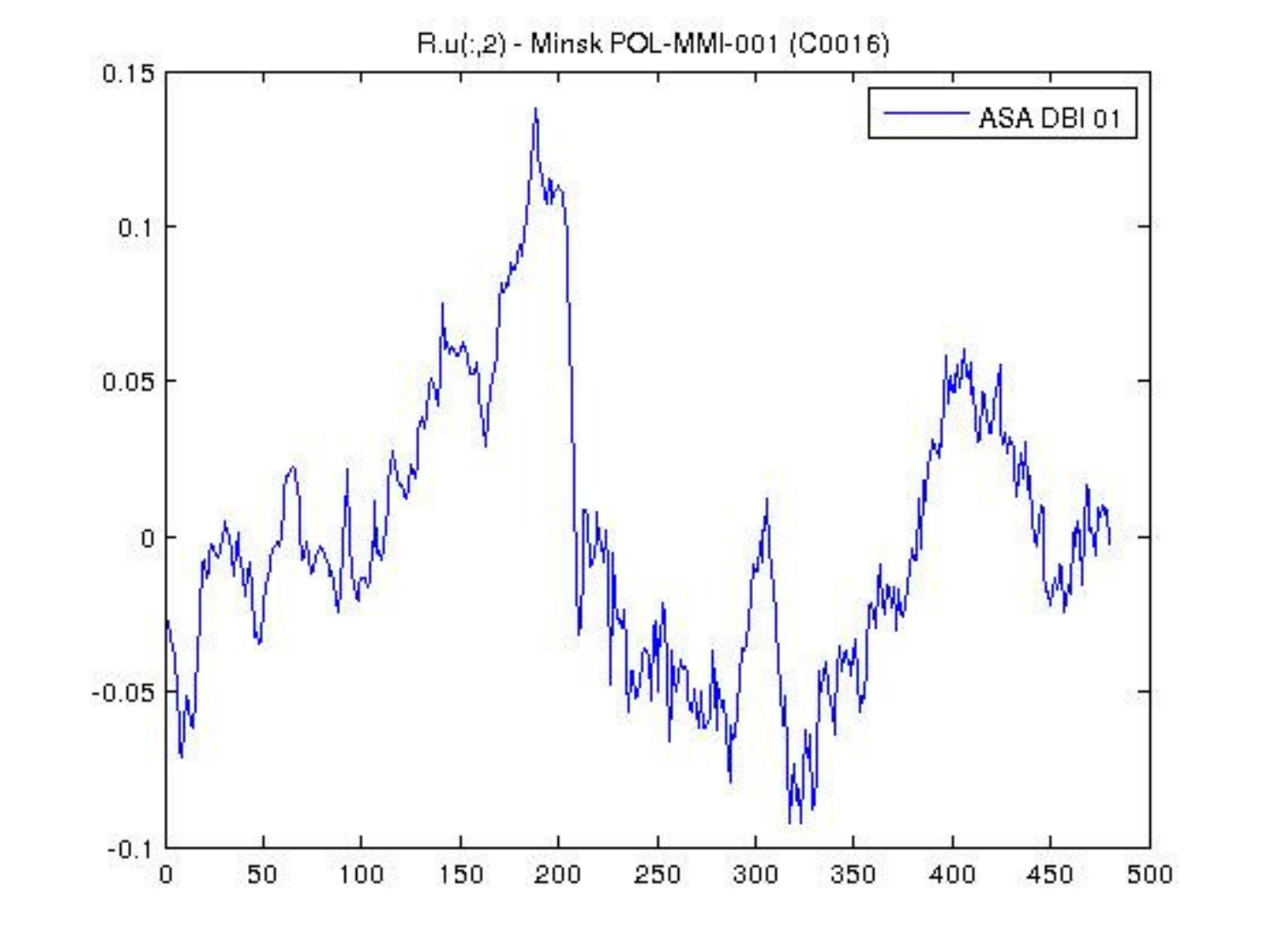}
\caption{$q_i$ vs $i$ (left) and ${\bf v}_2$ (right),
for a second signal from Mi\'{n}sk Mazowiecki}
\label{minsk2}
\end{figure}

Geostationary satellites are not completely stationary relative to stations
on the ground, moving north-south and east-west due to their orbital
inclination, eccentricity, and longitude drift. This leads to a 24-hour
variation in the signal strength at the receiving station, which can be
seen in the plots of $q_i$ in figures \ref{rugby1} and \ref{rugby2}.
It is also present in $q_i$ signals plotted in figures \ref{minsk1} and
\ref{minsk2}, though it is not as obvious in those plots. We use the
singular value decomposition to generate `eigensignals' for state vectors,
based on this 24-hour periodicity.

We consider $\mathrm{EIRP}$ data for $m$ days, with $n$ values per day,
${\bf E}=(E_1, E_2, \dots E_{N})$, where $N = mn$, and express the values as
a state vector ${\bf q}$ using equations \ref{ei-def}, \ref{pi-def}, and
\ref{qi-def}, then we define the Matrix ${\bf M}$ in terms of the elements of
${\bf q}$ to be
\begin{equation}
{\bf M} =
\left(
\begin{array}{cccc}
q_1          & q_2          & \ldots & q_n    \\
q_{n+1}      & q_{n+2}      & \ldots & q_{2n} \\
q_{2n+1}     & q_{2n+2}     & \ldots & q_{3n} \\
\vdots       & \vdots       &        & \vdots \\
q_{(m-1)n+1} & q_{(m-1)n+2} & \ldots & q_{mn}
\end{array}
\right) ,
\end{equation}
so that each row corresponds to data from one day.
Taking the singular value decomposition (SVD) of ${\bf M}$, we can write:
\begin{equation}
{\bf M} = {\bf U}{\bf S}{\bf V}^T
\end{equation}
where ${\bf U}$ is an $m \times m$ orthogonal matrix,
${\bf S}$ is an $m \times n$ diagonal matrix containing non-negative
real numbers, and ${\bf V}$ is an $n \times n$ orthogonal matrix.
The columns of ${\bf V}$ are called the right-singular vectors of ${\bf M}$,
and they are orthonormal and are the principal components of the rows of
${\bf M}$. We refer to as ${\bf v}_i$, and
in fact they are the eigenvectors of the covariance matrix
${\bf M}^T {\bf M}$, and hence we call them `eigensignals', and they are
characteristic for the 24-hour periods. In a celebrated paper~\cite{TurkP1991}
this technique was applied to the classification of human faces.
The scalars $\alpha_i$ are ordered so that $\alpha_1$ is the largest, and they
decrease with increasing $i$, so the ${\bf v}_i$ vectors with small values
of $i$ make the greatest contribution to the sum, and therefore to the signal.
The columns of ${\bf U}$ are the left-singular vectors of ${\bf M}$, and they
contain the information on the proportion of each eigensignal that is
present in the signals for each day.

The ${\bf v}_1$ vector picks out the dominant part of the 24-hour variation
in the signal, which turns out not to be very characteristic of the uplink
antenna, and to be rather similar for all carriers sharing the same downlink.
However, the ${\bf v}_2$ vector {\em is} characteristic of the uplink antenna,
and we therefore use ${\bf v}_2$ as the signature.
Figure \ref{rugby1} shows $q_i$ and ${\bf v}_2$
for a carrier transmitted from an antenna at a station in Rugby (England)
over the 31 days in December 2012, and figure \ref{rugby2} shows the
corresponding plots for another carrier transmitted from the same antenna in
Rugby during the same period. As can be seen, the ${\bf v}_2$ values are very
similar to each other. For comparison, figures \ref{minsk1} and \ref{minsk2}
show the corresponding plots for two carrier signals from an antenna in
Mi\'{n}sk Mazowiecki (Poland) during the same period, and again
the ${\bf v}_2$ values are very similar to each other, but quite different
to the ones for the signals from the antenna in Rugby.

\section{Quantifying the Similarity between Signatures}
\label{quantify-similarity}

The scalar product of two quantum state vectors ${\bf r}$ and ${\bf s}$
with real components is ${\mathbf r}^T {\mathbf s}$ and we can use this to
define a measure $D$ of the distance\footnote{In the complex case
$D({\mathbf r}, {\mathbf s}) = \sqrt{1 - |{\mathbf r}^\dagger {\mathbf s}|^2}$}
between the states,
\begin{equation}
D({\mathbf r}, {\mathbf s}) = \sqrt{1 - |{\mathbf r}^T {\mathbf s}|^2}
\end{equation}
which is zero when the vectors are identical, and is one
when they are maximally different (orthogonal).
Since the signatures (the ${\bf v}_2$ vectors from the SVD)
are orthonormal, they have a norm of one, and they are valid quantum state
vectors, and we use $D$ to calculate the distance between pairs of them, to
quantify their similarity. The distance for carriers from the same antenna
turns out to be lower on average than for carriers from different antennas.

% Note that $D$ can also be used to calculate distances between state vectors
% like $\mathbf{q}$, as defined in equation~\ref{q-def}, however the maximum
% distance between two such vectors is $\frac{1}{\sqrt{2}}$, because the $q_i$
% values are non-negative, which in turn is because the probability values
% $p_i$ cannot be negative. The components of the ${\bf v}_2$ vectors can be
% negative though, and it is possible for two of them to be separated by the
% maximum distance of $1$, which is an example of the benefits that can
% arise from representing signal data as quantum states.

\section{Performance Evaluation}
\label{performance-evaluation}

In order to quantitatively test this approach for identifying signals
we developed a statistical model based on histograms of distances
between carriers, and applied the model to carrier data that was monitored
in Dubai in December 2012, consisting of $53$ carriers from $32$ antennas,
which were relayed by the SESAT2 satellite. We initially analysed data from
the whole month, and then investigated the performance for shorter periods
of time.

\subsection{Statistical model and results for data from one month}

\begin{figure}[htb]
\begin{center}
\includegraphics[width=60mm]{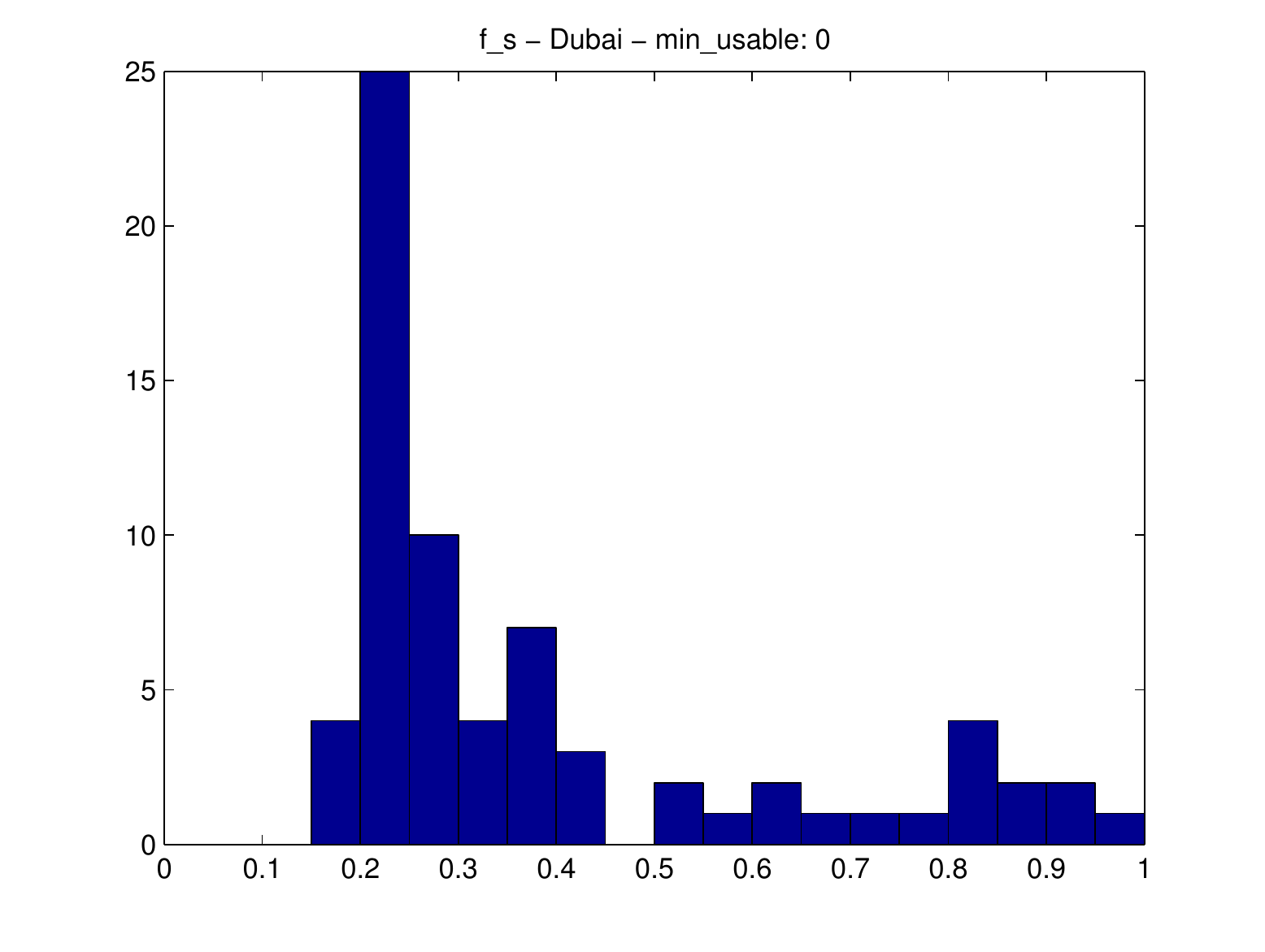}
\includegraphics[width=60mm]{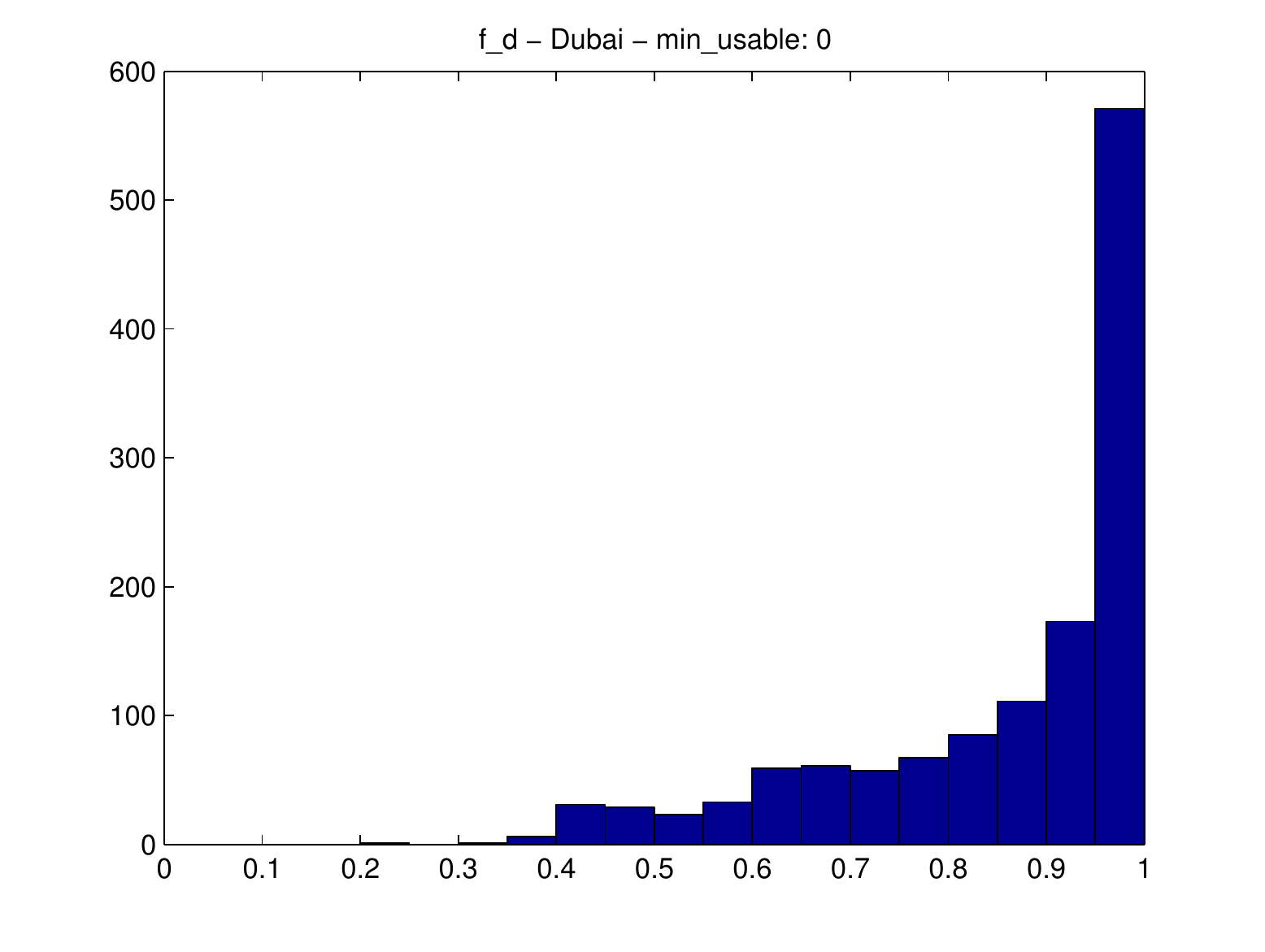}
\caption{$f_s(D)$ (left) and $f_d(D)$ (right) - Dubai}
\label{dubai-one-month}
\end{center}
\end{figure}

Figure \ref{dubai-one-month} shows the frequency distributions of distances
between pairs of different carriers, from the same antenna, $f_s(D)$, and from
different antennas, $f_d(D)$, using data for each carrier for the whole month.
For the 53 carriers there are $\binom{53}{2} = 53 \times 26 = 1378$ pairs,
70 of which were from the same antenna, and 1308 of which were from different
antennas. It can be seen that the distances between most carriers from
the same antenna are much lower than the distances between carriers from
different antennas, so fairly good separation can be obtained between
carrier pairs from the same antenna and pairs from different antennas.

\begin{figure}[htb]
\begin{center}
\includegraphics[width=60mm]{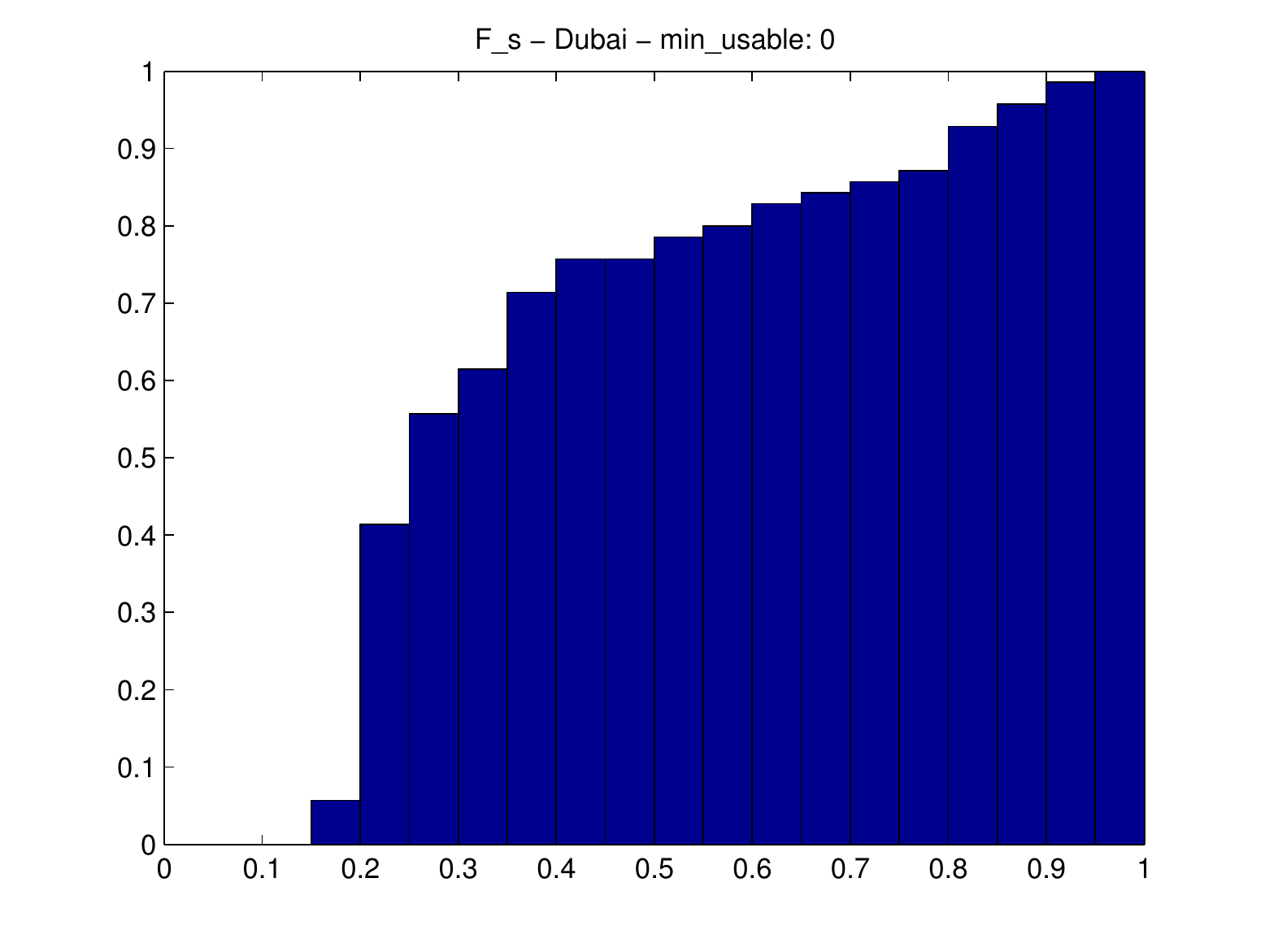}
\includegraphics[width=60mm]{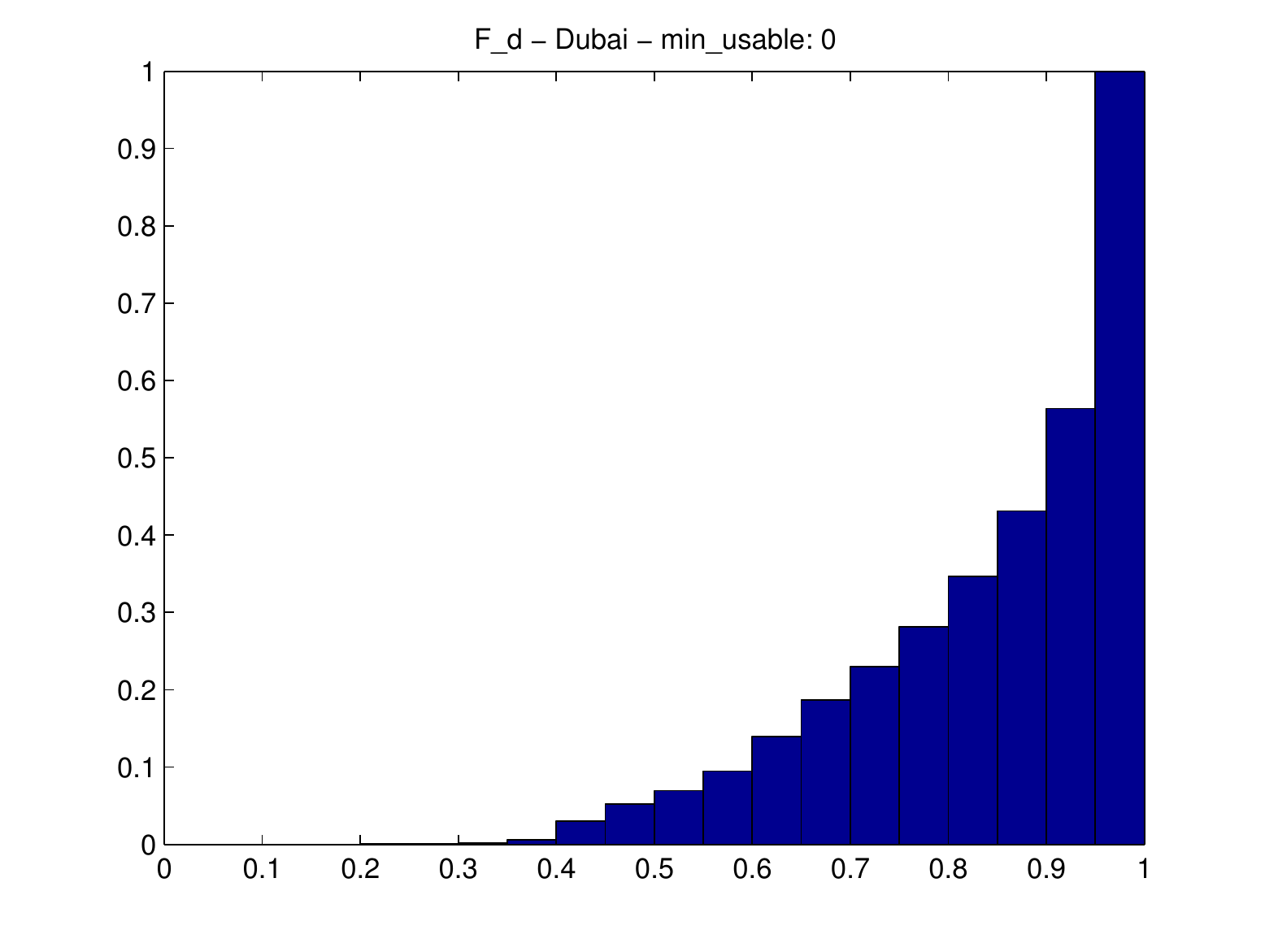}
\caption{$F_s(D)$ (left) and $F_d(D)$ (right) - Dubai}
\label{dubai-one-month-cum}
\end{center}
\end{figure}

The quality of the separation can be characterized using the corresponding
relative cumulative frequency distributions, for pairs of carriers from the
same antenna, $F_s(D)$, and from different antennas, $F_d(D)$, shown in
figure \ref{dubai-one-month-cum}. $F_s(D)$ is an estimate of
the probability that the distance between  a pair of carriers from the same
antenna is less than $D$. For example, the value of $F_s(0.2)$
is approximately $0.71$. Given the scenario that an interfering carrier is
coming from one of a number of known uplink antennas, but we don't know which
one, our approach for identifying the antenna is to calculate the distances
from the interferer to all the known carriers from the satellite that is
relaying the interferer, and to select those for which the distance $D$ is
less than a specified threshold $D_t$, and we refer to those carriers as being
in the `result set'. $F_s(D_t)$ is therefore an estimate of the probability
that a given carrier from a satellite is in the result set. Each carrier in
the result set that is from the same antenna as the interferer is a positive
identification of the source of the interferer, so we refer to these as
{\em positives}.

\subsubsection{Probability of at least one positive}

Typically several carriers are transmitted from the same antenna at the same
time, so the probability that at least one carrier from the same antenna as
the interferer is in the result set is bigger than $F_s(D_t)$, because
$F_s(D_t)$ is a probability estimate for a single pair, but on average there
is more than one of them. We call this probability $p_{id}$, because it is the
probability that at least one carrier has been correctly identified as coming
from the same antenna as the interferer. For each known carrier, the
probability that it is {\em not} in the result set is $1 - F_s(D_t)$, so for
an antenna with $k$ carriers the probability that none of them are in the
result set is $(1-F_s(D_t))^k$. If the number of antennas that have $k$
carriers is $n(k)$, then averaged across antennas, the probability that none
of the carriers are in the result set, which we call $\bar{p}_{id}$, is
\begin{equation}
\bar{p}_{id} = \frac{\sum_k n(k) (1-F_s(D_t))^k}{\sum_k n(k)} .
\end{equation}
Now, the sum in the denominator of this equation is equal to the total number
of antennas, which we call $N_a$ and the probability that {\em at least one}
carrier in the result set came from the same antenna as the interferer is
$1 - \bar{p}_{id}$, so
\begin{equation}
p_{id} = 1 - \frac{1}{N_a} \sum_k n(k) (1-F_s(D_t))^k .
\end{equation}
For the Dubai data in December 2012, 27 of the 32 antennas had only one
carrier, and the other five antennas had two, three, six (twice) and nine
carriers respectively, so   if we choose $D_t = 0.4$ for our threshold, for
which $F_s(D_t) = 0.714$, the probability that the antenna we are seeking
is in the result set is approximately $p_{id} = 0.76$.

\subsubsection{Expected number of positives}

We call the expected number of positives $n_i$, because they are carriers
that have been correctly {\em identified} as coming from the same antenna as
the interferer, and we call the average number of carriers per antenna $n_s$,
because it is the average number of carriers from the {\em same} antenna.
Only carriers from the same antenna as the interferer can contribute towards
$n_i$, and on average there will be $n_s$ of them. We can consider the
decision as to whether the distance of each of these carriers from that of the
interferer is less than $D_t$ as being independent trials, so so the
probability that a carrier from the same antenna as the interferer is in the
result set (the trial is a success) is equal to the number of successes $n_i$
divided by the number of trials $n_s$, but we know that this probability is
given by $F_s(D_t)$, so we have $F_s(D_t) = n_i/n_s$, which gives an estimate
for $n_i$ of
\begin{equation}
n_i = n_s F_s(D_t) .
\label{eq_n_i}
\end{equation}
A more detailed analysis takes into account the distribution of the number
of carriers per antenna.
We can consider the decisions as to whether the distances between
each carrier from the antenna and the interferer are less than $D_t$ as being
independent trials, so the probability of obtaining $k$ positives from an
antenna with $K$ known carriers, which we call $P(k)$, will follow a binomial
distribution, and if we let $p = F_s(D_t)$ and $q = 1-p$, then
\begin{equation}
P(k) =  \binom{K}{k}p^kq^{K-k} .
\end{equation}
The expected value of $k$ is known to be
\begin{equation}
\bar{k} = \sum_k kP(k) = Kp ,
\end{equation}
and given the distribution $n(K)$, and averaging over $K$, the expected
number of positives is
\begin{equation}
n_i = \langle\bar{k}\rangle = \frac{p}{N_a} \sum_K n(K)K = \bar{K}p ,
\end{equation}
where $\bar{K} = n_s$, the average number of carriers from the same
antenna, so we obtain the same result as equation~\ref{eq_n_i}.

For the Dubai data in December 2012 the value of $n_s$ was approximately
$1.66$ and $F_s(0.4)$ was $0.71$, so $n_i$ is therefore approximately $1.2$.

\subsubsection{Expected number of false positives}

In general the result set will also contain some carriers from antennas
other than the one that we are trying to identify, and the number of such
{\em false positives}, which we call $n_f$, can be estimated using $F_d(D)$,
which is an estimate of the probability that the distance between a pair of
carriers from different antennas is less than $D$. A similar argument to
the one that led to equation \ref{eq_n_i} gives
\begin{equation}
n_f = n_d F_d(D_t),
\end{equation}
where $n_d$ is the number of carriers from different antennas to the one
that the interferer originated from. We call the number of carriers relayed by
the satellite is $N_s$, and we know that of these, on average $n_s$ of them
are from one particular antenna, so the average the number that are from
other antennas is
\begin{equation}
n_d = N_s - n_s ,
\label{eq_n_d}
\end{equation}
so we have
\begin{equation}
n_f = (N_s - n_s) F_d(D_t),
\label{eq_n_f}
\end{equation}

For this example case, 53 carriers were relayed by the satellite, so
$N_s = 53$, and given our example threshold of $D_t = 0.4$, the value of
$F_d(0.4)$ is approximately $0.0061$, so we have an estimated number of
$n_f = n_d F_d(D_t) = 51.34 \times 0.0061$ false positives, which is
approximately $0.31$. Note that a slightly higher value of $D_t$, say
$D_t = 0.5$, would give a higher expected number of positives,
of $n_i = 1.25$, but also a much higher number of false positives,
$n_f = 2.7$.

Note also that we need not restrict our data to carriers relayed by the same
satellite as the interferer, since other satellites may also relay carriers
from the antenna that is the source of the interferer, but it was found that
for the data at our disposal, including such carriers increased the false
positive rate significantly, with little or no increase in the number of
positives.

\subsubsection{Probability of one or more false positives}

For a carrier from an antenna with $K$ carriers there are $N_s -K$ possible
comparisons with carriers from the other antennas, and if we now let
$p = F_d(D_t)$ then the probability of one or more false positives, which we
call $p_f^K$, as it applies to an antenna with $K$ carriers, is
\begin{equation}
p_f^K = 1 - (1 - p)^{(N_s-K)} .
\end{equation}
Provided $p$ is small, this approximates to
\begin{equation}
p_f^K \simeq (N_s-K)p ,
\label{eq_p_f}
\end{equation}
which happens to also equal the expected number of false positives for that
antenna, which we call $n_f^K$. Provided $N_s$ is large and $K$ is small,
(\ref{eq_p_f}) is not a strong function of K, and with this assumption
\begin{equation}
p_f^K = n_f^K \simeq N_s p \simeq \langle p_f^K \rangle = \langle n_f^K \rangle ,
\end{equation}
and this would be the expected result averaged over all the antennas, which
we call $p_f = \langle p_f^K \rangle$, and we have
\begin{equation}
p_f \simeq N_s p = N_s F_d(D_t),
\end{equation}
For the Dubai data in December 2012 the probability of a false positive is
therefore approximately $p_f = 0.32$.

\subsection{Results for data from two days}

\begin{figure}[htb]
\begin{center}
\includegraphics[width=60mm]{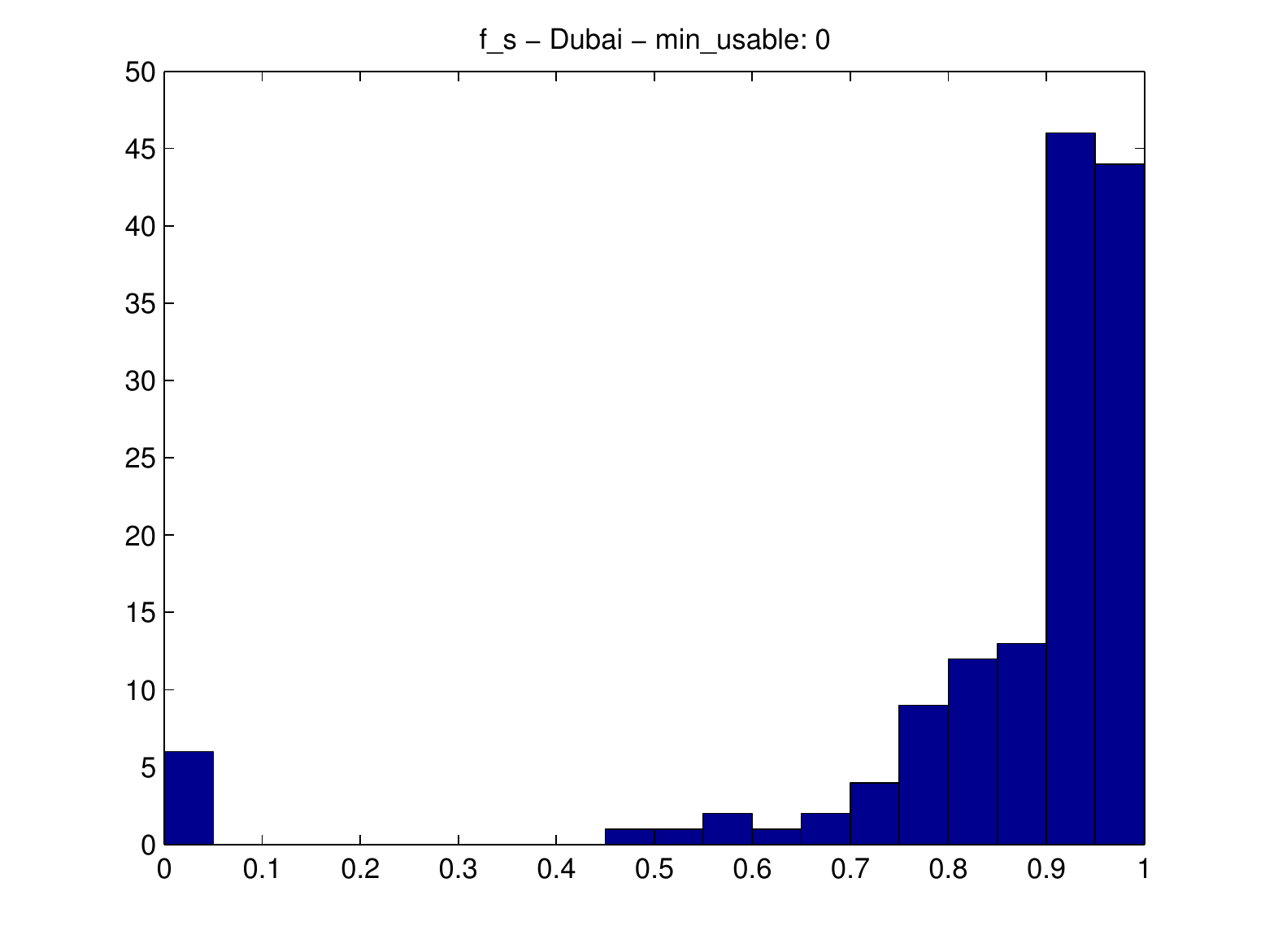}
\includegraphics[width=60mm]{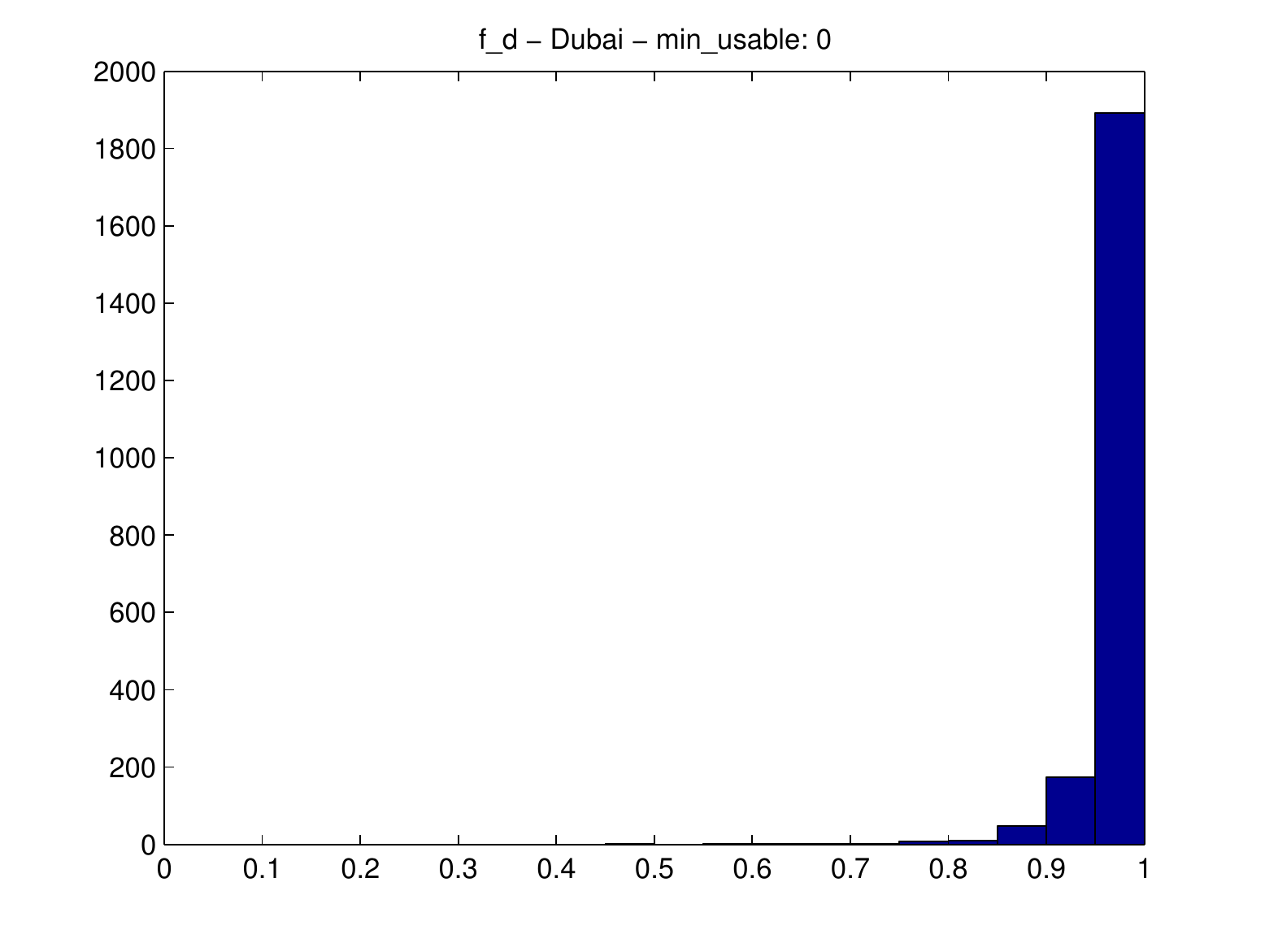}
\caption{$f_s(D)$ (left) and $f_d(D)$ (right) - Dubai, 1-2 Dec. 2012}
\label{dub-2d-dec1}
\end{center}
\end{figure}

\begin{figure}[htb]
\begin{center}
\includegraphics[width=60mm]{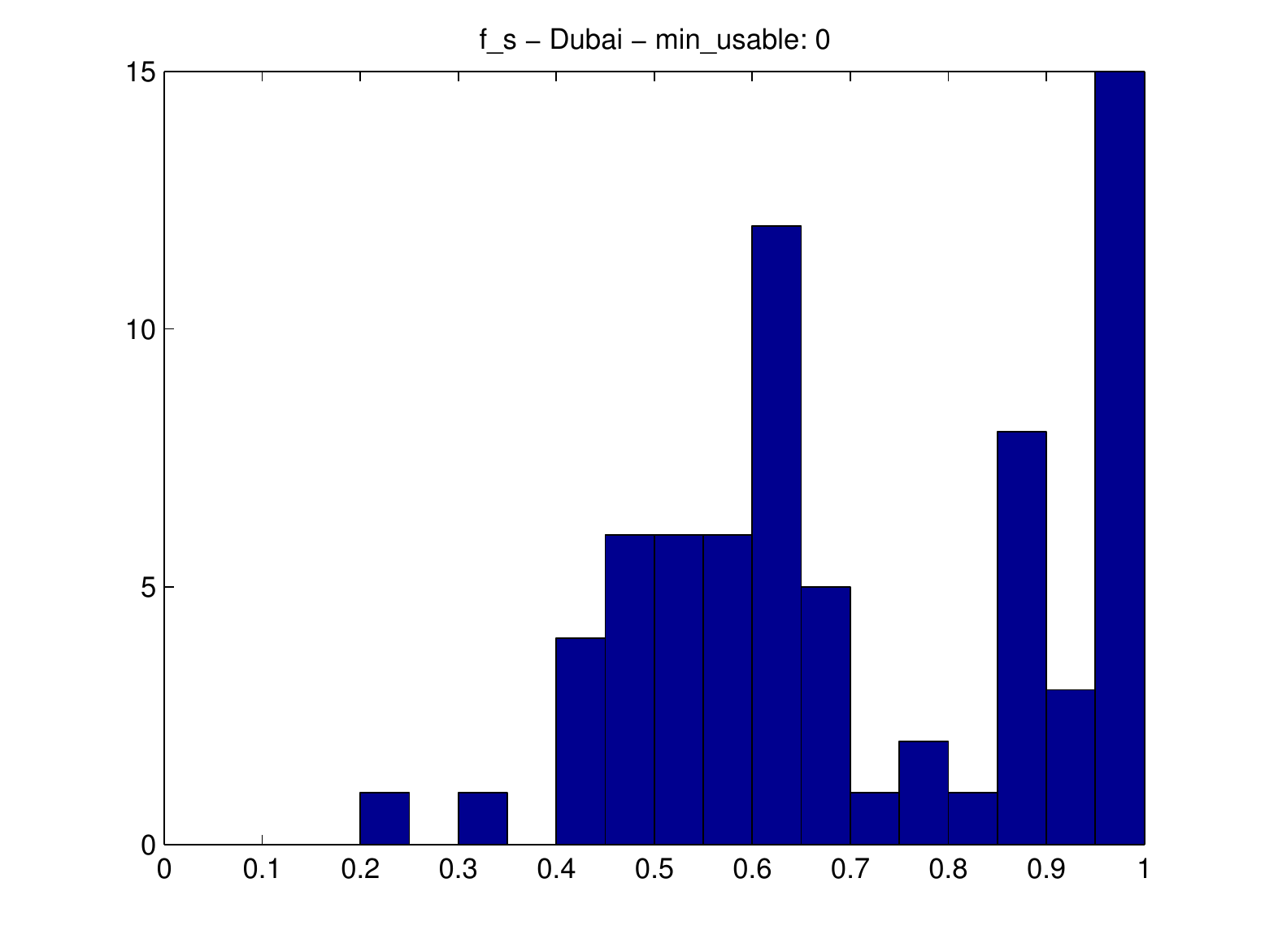}
\includegraphics[width=60mm]{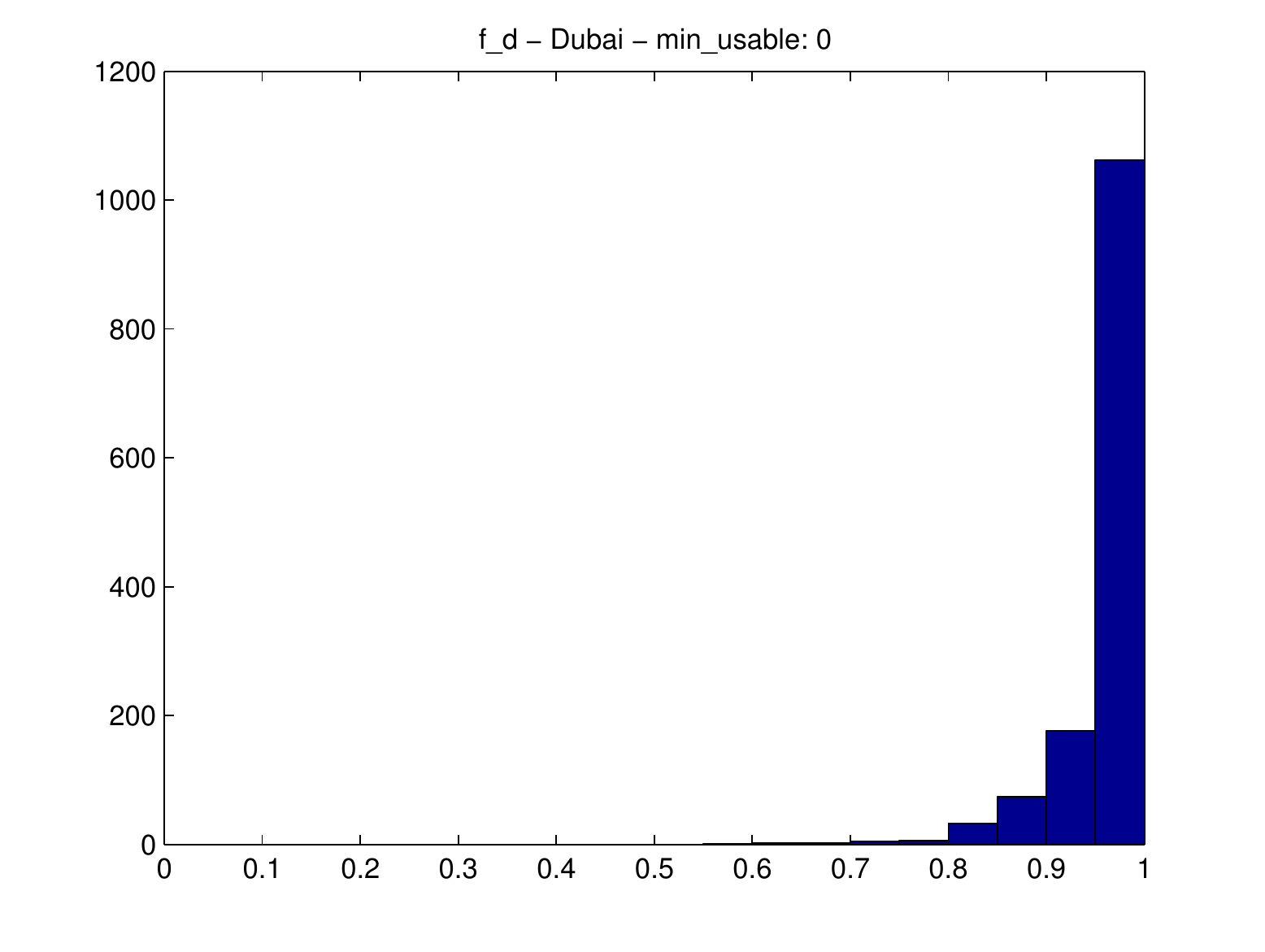}
\caption{$f_s(D)$ (left) and $f_d(D)$ (right) - Dubai, 15-16 Dec. 2012}
\label{dub-2d-dec15}
\end{center}
\end{figure}

So far we have presented data from a period of one
month, but we also tried analysing data from shorter periods of time, and
found that results with $n_i > n_f$ could be obtained for periods right down
to two days, which was the minimum amount of data necessary for the algorithm
to work in its original form. However, we found that the results were better
for some two-day periods than for others. This can be seen from figures
\ref{dub-2d-dec1} and \ref{dub-2d-dec15}, which show
the frequency distributions of distances between pairs of carriers from the
same antenna, $f_s(D)$, and from different antennas, $f_d(D)$, for data
monitored by Dubai for two different two-day periods in December 2012.
Note that the number of counts in figure 7~(left) is roughly twice that of
figure 8~(left), which is because not all carriers were present for the whole
month, and for each period, only carriers were considered that were present
for the whole of the period. The data in figure 7 is for pairs taken from 68
carriers, and in figure 8 it is from 54 carriers. The average number of
carriers per antenna was also higher for the data in figure 7, which further
boosted the number of pairs.

The best separation is shown by the data from 15-16 December 2012, for which a
choice of $D_t = 0.7$ gives $p_{id} = 0.64$, $n_i = 0.97$, $n_f = 0.19$, and
$p_f = 0.20$.

\subsection{Results for data from less than two days}

The algorithm was designed to require a minimum
of two days of data (two periods of 24 hours each), as it compared one
day with the next, in order to remove the 24 hour variation that is present
in all carriers, and all of the results presented in the previous sections
used that version of the algorithm. We then decided to investigate whether
the algorithm could still identify characteristic features of carriers
if it was modified to use periods of less than 24 hours, which would mean
that the 24 hour variation was not removed. The modified algorithm still
compares data from equal-length periods, and there must be at least two
of them, but they can be of arbitrary length, and in particular, less
than 24 hours.

\begin{figure}[htb]
\begin{center}
\includegraphics[width=60mm]{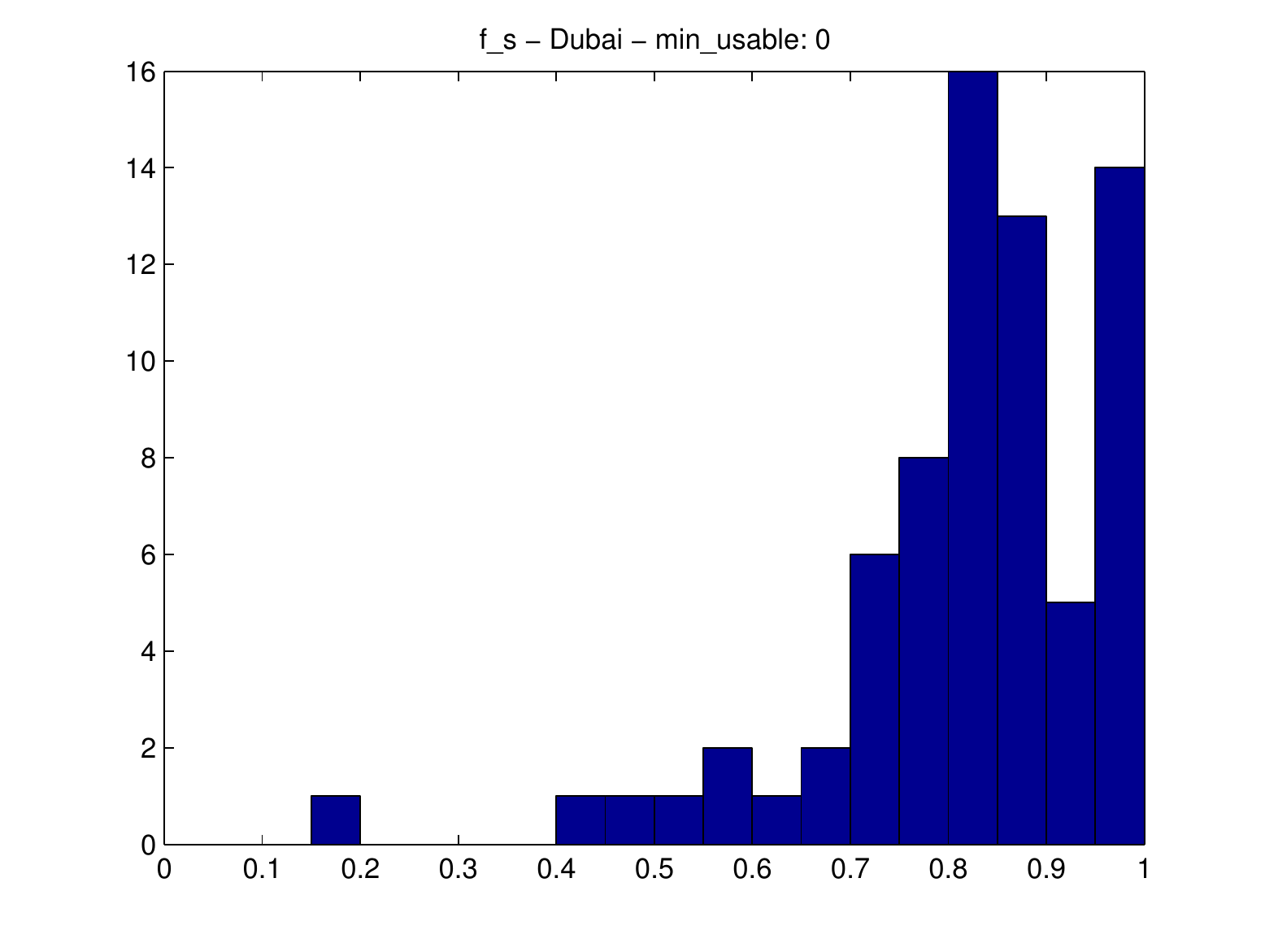}
\includegraphics[width=60mm]{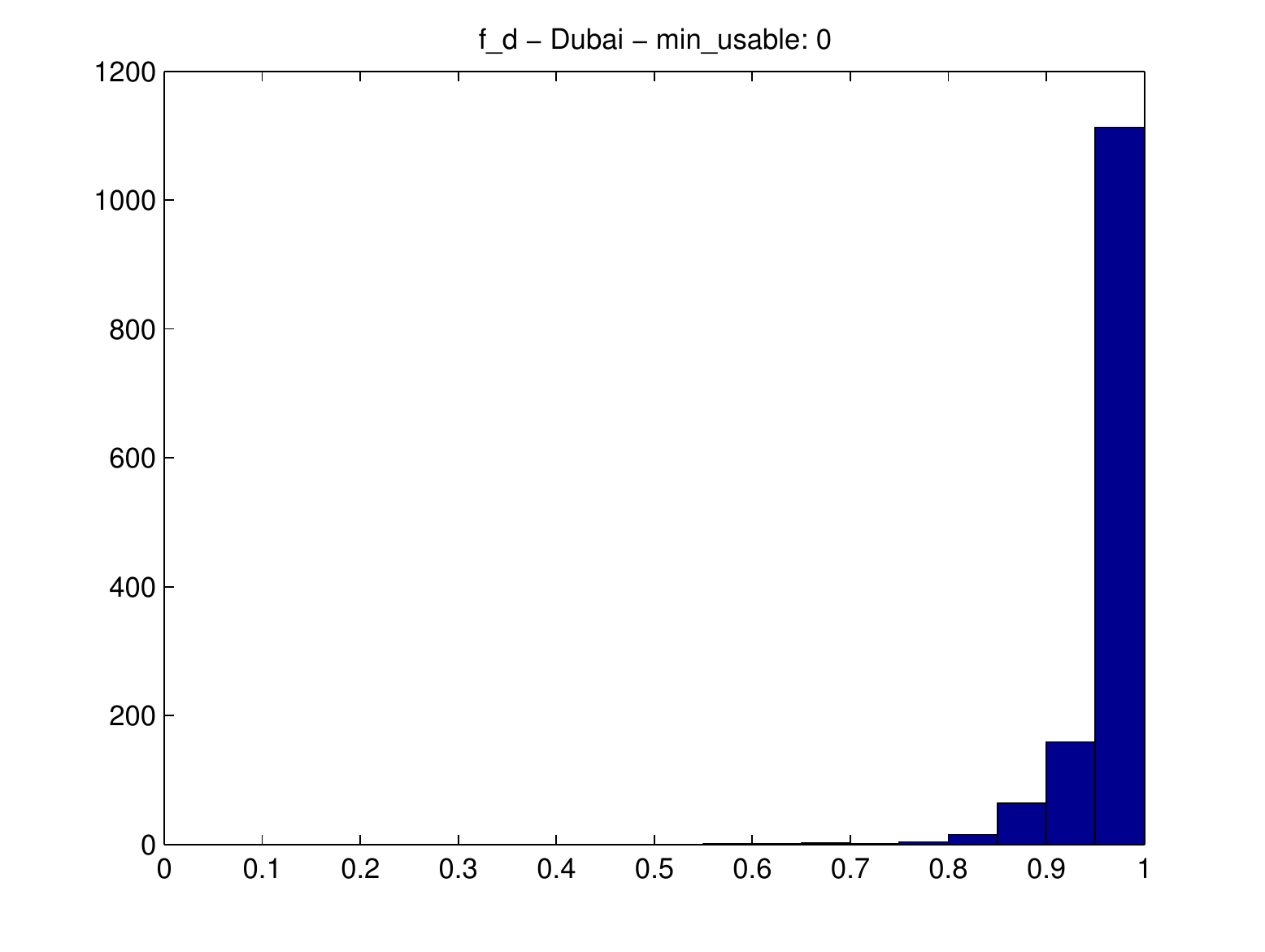}
\caption{$f_s(D)$ (left) and $f_d(D)$ (right) - Dubai, 15 Dec. 2012}
\label{dub-1d-dec15}
\end{center}
\end{figure}

Figure \ref{dub-1d-dec15} shows the frequency distributions of distances
between pairs of carriers from the same antenna, $f_s(D)$, and from different
antennas, $f_d(D)$, for one day's data, monitored at Dubai for December
15th 2012, split into two 12 hour periods. It can be seen that there is
some separation between carrier pairs from the same antenna and pairs from
different antennas, and choosing $D_t = 0.85$ gives $p_{id} = 0.62$,
$n_i = 0.93$, $n_f = 0.92$, and $p_f = 0.95$.

\begin{figure}[htb]
\begin{center}
\includegraphics[width=60mm]{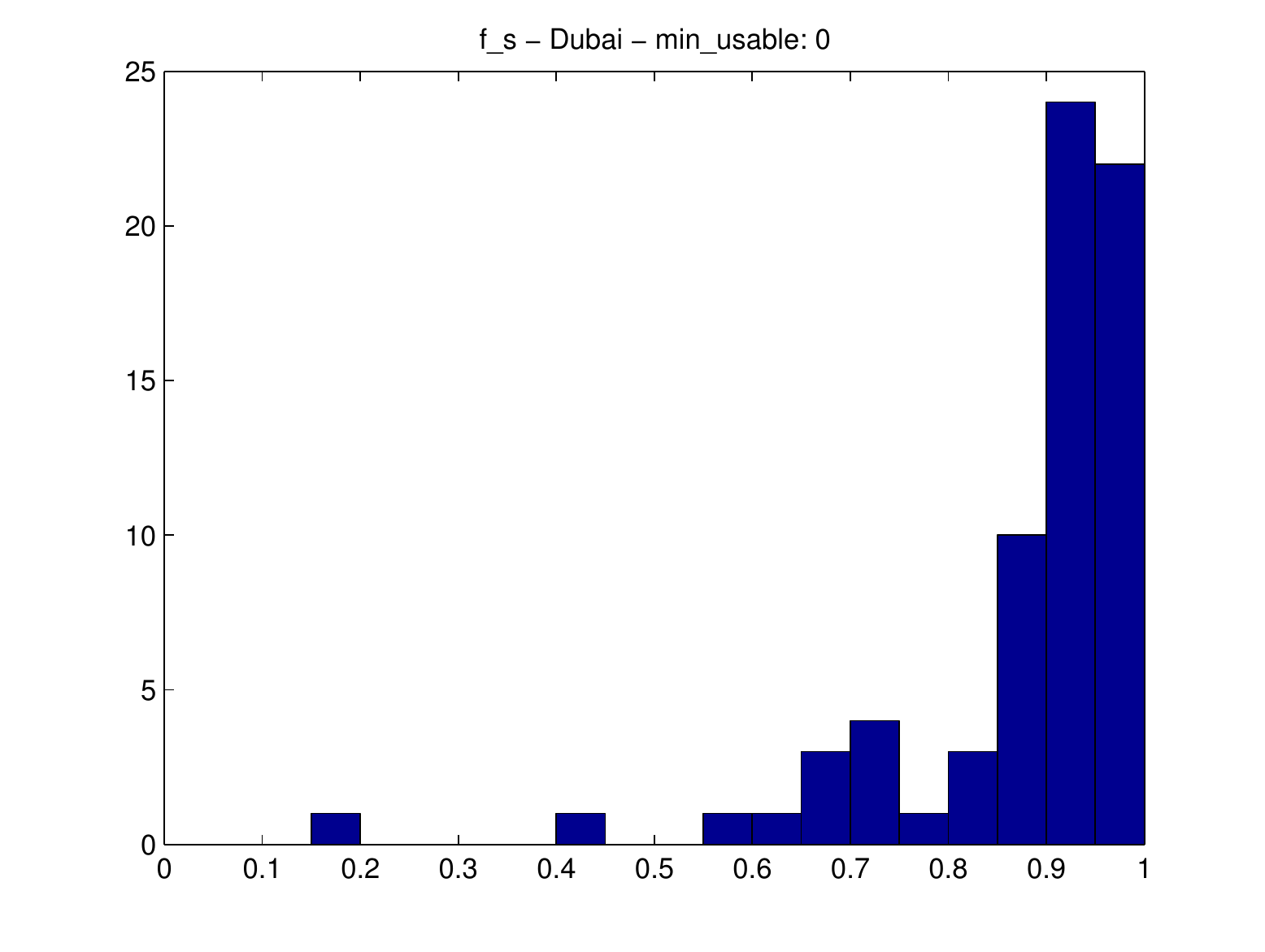}
\includegraphics[width=60mm]{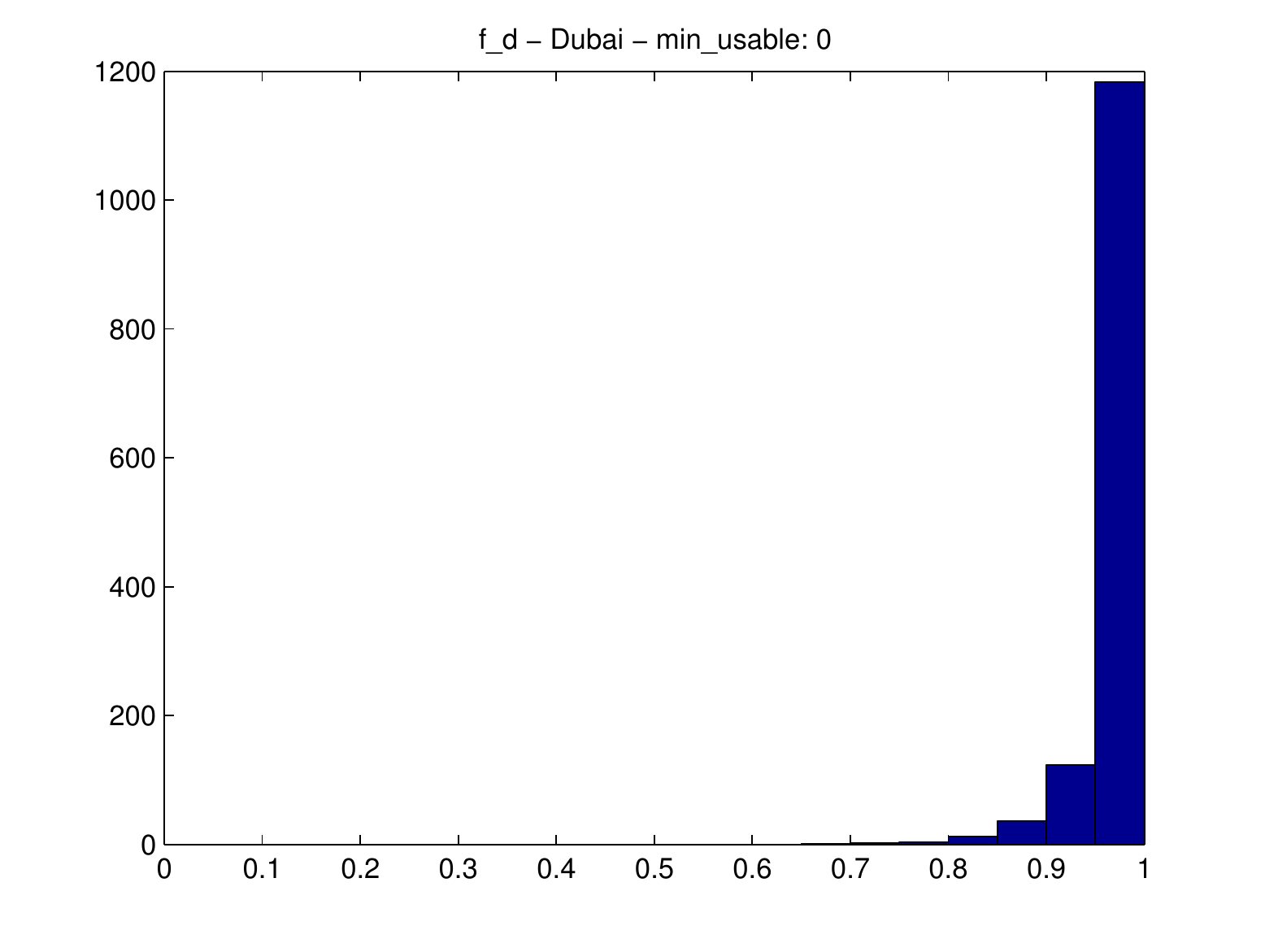}
\caption{$f_s(D)$ (left) and $f_d(D)$ (right) - 0:00--12:00, Dubai, 15 Dec. 2012}
\label{dub-0d5-1-dec15}
\end{center}
\end{figure}

Figure \ref{dub-0d5-1-dec15} shows the frequency distributions of distances
between pairs of carriers from the same antenna, $f_s(D)$, and from different
antennas, $f_d(D)$, for half a day's data, monitored at Dubai for the first
12 hours of December 15th 2012, split into two 6 hour periods. Again it can be
seen that there is some separation between carrier pairs from the same antenna
and pairs from different antennas, and choosing $D_t = 0.85$ gives $p_{id} = 0.33$,
$n_i = 0.33$, $n_f = 0.69$, and $p_f = 0.71$.

It is encouraging that the algorithm works at all for less than one day's
data, and although the results presented here show that it works less well
for shorter periods, this is based on a fixed sampling rate of $\mathrm{EIRP}$
values, meaning that the shorter periods contain less data. As long as the
signal contains enough structure, a higher sampling rate should give better
results.

\section{Conclusion}
\label{conclusion}

We have described a method for identifying the source of a satellite
interferer using a single satellite, which relies on the variation with time
of the strength of carrier signals measured at the downlink station. The
method uses a quantum-inspired algorithm to compute a signature for each
carrier, and a distance between the signature for an interfering carrier and
the signatures of all known carriers being relayed by the same satellite.
As a proof of concept we have presented a simple statistical model to
estimate the probability of successful
identification of the source of an interferer, the expected number of
carriers correctly identified to have originated from the interfering
transmitter, the expected number of false positives, and the probability of
one or more false positives, and we have used the model
to evaluate the performance of the technique using measured data for a sample
of 53 carriers relayed by one satellite. We presented results using data from
one month, and also for two days and less. In its original
form the algorithm was designed to work with a minimum of two days' of data,
and we found that the results were better for some two-day periods than
others, but that in some cases successful identification was possible. We also
modified the algorithm to operate on less than two days' of data, and we
found that the results were less good, but that positive identification
was still possible in some cases. However, the results were based on a fixed
sampling rate, meaning that the shorter periods contained less data, and
a higher sampling rate should give better results.

\section*{Acknowledgements}

This work was supported by the Austrian Research Promotion Agency (FFG)
and the European Space Agency (ESA). We are indebted to Alexander Ploner for
invaluable advice on the statistical analysis of the measured data, and to
Thomas Zemen and Piotr Gawron for comments that greatly improved the
manuscript.

\bibliographystyle{unsrt}
\bibliography{refs}

\end{document}